\documentclass[12pt,a4paper]{article}
\pdfoutput=1

\usepackage{amssymb}
\usepackage{amsmath}
\usepackage{amsfonts}
\usepackage{amsthm}
\usepackage{mathrsfs}
\usepackage{booktabs} 
\usepackage{amsmath} 
\usepackage{tabularx}  
\usepackage{enumitem}
\usepackage{setspace}
\usepackage{color}
\usepackage[pdftex]{graphicx}
\usepackage{authblk}
\usepackage{subfigure}
\usepackage{float}
\usepackage[utf8]{inputenc}
\usepackage{subfigure}  			
\usepackage{adjustbox}
\usepackage[font=small]{caption}	
\usepackage[compress]{cite}         
\usepackage{float}
\usepackage{url}

\usepackage{datetime}

\usepackage{pgfplots}   			
	\pgfplotsset{width=10cm}
	\usepackage{bbm}
	\usepackage{tensor}
	\usepackage{slashed}

\numberwithin{equation}{section}

\theoremstyle{plain}

\usepackage{color}

\newdateformat{daymonthyear}{\THEDAY \, \monthname[\THEMONTH] \THEYEAR}
\newdateformat{monthyear}{\monthname[\THEMONTH] \THEYEAR}

\pgfplotsset{compat=1.17}

\begin{document}

\title{Enhancement and Suppression of Decay Rates in an Accelerated Fermionic Cavity Coupled to a Massive Field}


\author{Vladimir Toussaint\thanks{{\tt vladimir.toussaint@nottingham.edu.cn}} }
\affil{School of Mathematical Sciences,\\ 
University of Nottingham Ningbo China,\\
Ningbo 315100, PR China}

 \date{\daymonthyear\today}

\maketitle
\begin{abstract}
We study a (1+1)-dimensional model in which a massless Dirac field, initially in an excited state inside a uniformly accelerated cavity, decays to its ground state, accompanied by the excitation of an external massive Dirac field of mass $M$, through a local coupling confined to the physical extent of the cavity.
The confinement mechanism is modeled via MIT bag boundary conditions and their probabilistic extensions, which depend on a boundary angle $\theta \in[0,2\pi)$ and $s\in(0,1)$.  For intermediate-sized cavities 
($a l \sim c^2$) with light external massive Dirac field ($Mc^2 \ll \hbar a/c$),  we demonstrate that the total long-time asymptotic decay rate factorizes as $
\Gamma_{\text{acc}}/ \Gamma_{\text{in}} \sim F_g F_T
$ with $\Gamma_{\text{in}}$ the inertial decay rate. Here, $F_g=\frac{al/c^2}{\ln(1 + al/c^2)}$ is a geometric factor, and $F_T= (1 + e^{-2\pi\beta})^{-1}$ is the thermal stimulation factor from the Unruh bath ($\beta = \frac{\Omega_1 c}{a} =\frac{(1+s)\pi}{\ln(1 + a l/c^2)}$). Crucially, in this regime, the thermal factor $F_T$ remains approximately unity for all admissible boundary conditions, while the geometric factor $\frac{a l/c^2}{\ln(1 + a l/c^2)}$ produces measurable enhancements up to 26\% for realistic parameters ($a=10^{20}$ m/s$^2$, $l=500~\mu$m), and represents
 a measurable signature accessible through quantum simulation platforms. In contrast, for heavy external fermionic fields (such as the electron field), the condition $M c^2 \gg \hbar a / c$ is satisfied at all achievable accelerations, placing the system in a regime of exponential suppression, $\Gamma_{\text{acc}}/\Gamma_{\text{in}} \sim \exp(-2 M c^2 / (\hbar a/c))$, for all cavity sizes. This suggests that, within this specific model, the observable decay signatures of acceleration are strongly suppressed, rendering Unruh-induced enhancements highly improbable in experiments involving heavy fermions under these conditions. 
\end{abstract}

\singlespacing

\newpage 

\section{Introduction\label{sec:intro}}

Quantum field theory in curved spacetime reveals deep connections between acceleration, gravity, and thermodynamics. A cornerstone of this framework is the Unruh effect \cite{Unruh1976,Davies1975,Fulling:1972md}, where a uniformly accelerated observer perceives the Minkowski vacuum as a thermal bath at temperature $T_{\text{Unruh}} = a/2\pi$. This effect is fundamental to our understanding of quantum phenomena in non-inertial frames and is intimately connected to black hole thermodynamics via the Hawking effect \cite{Hawking1975}. Despite its conceptual importance and decades of theoretical study, the Unruh effect has eluded direct experimental confirmation. Its verification remains a central challenge, requiring the detection of faint thermal signatures under extreme accelerations while distinguishing genuine Unruh radiation from alternative explanations \cite{Unruh:2014hua, Brenna:2015fga}.

Traditional detection approaches have focused on the response of individual accelerated particle detectors \cite{DeWitt1980, Louko2008}, but the required accelerations are often prohibitively high. Recent proposals have sought amplification through collective behavior or structured environments \cite{Retzker2005, Bruschi:2013yeu}. Cavity quantum electrodynamics (QED) settings offer a promising alternative \cite{Crispino2008,Hu:2004zu}, where field confinement enhances matter interactions and enables sensitive probes of vacuum thermality through discrete mode spectra.

This work examines how a massive external field and cavity geometry shape the decay behavior of an accelerated quantum system. The results reveal mechanisms of exponential suppression and geometric enhancement, providing insight into non-inertial effects and possible applications in quantum simulation.

Specifically, we analyze a uniformly accelerated cavity QED setup where a massless Dirac field confined within the cavity is coupled to an external massive Dirac field through an interaction that is spatially localized to the cavity’s region. Our analysis reveals that for heavy external fermionic fields with mass $M$, the condition $Mc^2 \gg \hbar a/c$ holds for all achievable accelerations, placing such systems in an exponentially suppressed regime that may hinder the observation of Unruh-induced enhancements in experiments involving heavy fermionic particles within our framework. In contrast, for a light external Dirac field ($Mc^2 \ll \hbar a/c$) and for intermediate-sized cavities $al\sim c^2$, we identify a measurable geometric enhancement arising from kinematic constraints that is more dominant than any Unruh-induced decay modification tied to the thermal bath.

The fermionic nature of both fields is crucial, as their statistics and boundary conditions at the cavity walls fundamentally alter mode structures and their overlap. We consider both MIT bag boundary conditions enforcing strict confinement \cite{MIT-bag-original} and more general probabilistic conditions parametrized by an angle $\theta \in[0,\pi/2)$ and a parameter $s\in(0,1)$ \cite{Friis:2011yd,Friis:2013eva}, allowing continuous tuning of field behavior at boundaries.

We study the decay rate of an excited state of the confined massless Dirac field—a direct probe of vacuum fluctuations experienced by a co-accelerated observer. While prior work has focused on scalar fields \cite{Lorek} or unconfined fermionic fields \cite{Vanzella:2001ec, Crispino2008}, few studies address combined effects of cavity geometry and fermionic statistics. Those examining fermions in cavities \cite{Friis:2011yd} have emphasized quantum information protocols like entanglement harvesting \cite{Friis:2013eva} rather than decay dynamics.

Our analysis fills this gap and reveals a robust, experimentally viable signature. The decay rate exhibits distinct behavior governed by two dimensionless parameters: the cavity size relative to acceleration scale ($al$) and the field mass relative to acceleration ($M/a$). For light external fields ($M/a \ll 1$), the rate transitions from inertial-like behavior ($\Gamma_{\text{acc}} \sim \Gamma_{\text{in}}$) for small cavities ($al \ll 1$) to geometric enhancement ($\Gamma_{\text{acc}} \sim \Gamma_{\text{in}}\frac{al}{\ln(1+al)}$) for intermediate cavities ($al \sim 1$), reaching up to 26\% enhancement for realistic parameters. For heavy external fields ($M/a \gg 1$), the rate is exponentially suppressed ($\Gamma_{\text{acc}}/\Gamma_{\text{in}} \sim e^{-2M/a}$) regardless of cavity size.

This comprehensive study provides a unified picture of acceleration effects in confined systems and identifies quantum simulation as a promising path forward. The geometric enhancement offers a measurable signature accessible through platforms like superconducting circuits and optomechanical systems.

The paper is organized as follows: In Sec.~\ref{sec:model}, we introduce our fermionic QED cavity model. Sections~\ref{sec:MassiveDiracField} and~\ref{sec:Massless-DField} review quantization of the external massive field in Minkowski spacetime and the confined massless field within an inertial cavity under various boundary conditions. Section~\ref{sec:intertial_decay} derives the inertial decay rate. Section~\ref{sec:rindler_quantization} covers Rindler quantization for both fields in a uniformly accelerated cavity. Section~\ref{sec:accelerated_decay} derives the accelerated decay rate. Section~\ref{sec:results} presents results across parameter regimes, highlighting geometric enhancement and exponential suppression. Section~\ref{sec:conclusion} discusses experimental prospects. Appendices provide details of asymptotic analyses for mode overlap integrals.

We use natural units ($\hbar = c = \kappa_B = 1$) throughout and work in (1+1)-dimensional Minkowski spacetime with signature $(+-)$. Limitations of this dimensionality are discussed in the conclusion.

\section{Model: Fermionic Fields in an Accelerated Cavity}
\label{sec:model}

To probe quantum decay dynamics in accelerated cavities, we introduce a minimal model that captures the essential physics of the Unruh effect. This model couples a confined, massless Dirac field (the ``clock") to an external, unconfined massive Dirac field, with interactions mediated by a Hermitian Hamiltonian. We first define the spacetime geometry, then quantize both fields, and finally impose boundary conditions to characterize the cavity system.

Consider a $(1+1)$-dimensional cavity of proper length $l$ in the lab frame at $t=0$. For an inertial observer, the lab frame is described by pseudo-Cartesian coordinates $(t, x)$ with the Minkowski metric:  
\begin{align}\label{eq:labframe-metric}
ds^2 = dt^2 - dx^2,
\end{align}  
where the cavity walls are fixed at $x = x_-$ (left wall) and $x = x_+ = x_- + l$ (right wall). These walls are rigidly attached to the timelike Killing vector $\partial_t$, ensuring staticity in the inertial frame.  

The ideal clock is modeled as a massless Dirac spinor field $\chi(t, x)$ confined to the cavity ($x \in [x_-, x_+]$), coupled to an external massive Dirac spinor field $\psi(t, x)$ (unconfined, $x \in \mathbb{R}$) with rest mass $M > 0$. The cavity walls are transparent to $\psi$, allowing coupling between the two fields.  

The interaction Hamiltonian, responsible for mediating energy exchange between $\chi$ and $\psi$, is:  
\begin{align}\label{eq:Hint_Coupling}
H_{\text{int}} = g \int_{x_-}^{x_+} dx \left( \bar{\chi}\psi + \bar{\psi}\chi \right),
\end{align}  
where $g$ is a small coupling constant with dimension $[g]=[M]$ in natural units ($\hbar=c=1$), ensuring $H_{\text{int}}$ has the correct dimension of energy $([M])$ in ($1+1$) dimensions. Here, we have $\bar{\chi} = \chi^\dagger \beta$, $\bar{\psi} = \psi^\dagger \beta$ (with $\beta$ defined in Section~\ref{sec:MassiveDiracField}). The Hermiticity of $H_{\text{int}}$ is ensured by the sum of the two terms, which guarantees real eigenvalues for physical observables.

\subsection{Quantization of the Massive Dirac Field}  
\label{sec:MassiveDiracField}

In the lab frame ($\eta^{\mu\nu} = \text{diag}(1, -1)$), the dynamics of the external massive spinor $\psi(t, x)$ are governed by the Dirac equation:  
\begin{align}\label{eq:2D-Dirac-eq}
i\partial_t \psi(t, x) = H_D \psi(t, x), \quad H_D = -i\alpha \partial_x + M\beta,
\end{align}  
where $H_D$ is the Dirac Hamiltonian, and $\alpha, \beta$ are $2 \times 2$ Hermitian matrices satisfying the Clifford algebra:  
\begin{align}
\{\alpha, \beta\} = 0, \quad \alpha^2 = \beta^2 = \mathbb{I}_2.
\end{align}  
These matrices are constructed from the gamma matrices $\gamma^\mu$ ($\mu = 0, 1$) via $\alpha = \gamma^0 \gamma^1$ and $\beta = \gamma^0$, with $\{\gamma^\mu, \gamma^\nu\} = 2\eta^{\mu\nu}\mathbb{I}_2$. Hermiticity of $H_D$ requires $\alpha^\dagger = \alpha$ and $\beta^\dagger = \beta$, consistent with $\gamma^{0\dagger} = \gamma^0$ and $\gamma^{1\dagger} = -\gamma^1$.

Following Ref.~\cite{Friis:2011yd}, we use a real two-component orthonormal spinor basis $\{U_+, U_-\}$ satisfying:  
\begin{subequations}\label{eq:real-spin-basis}
\begin{align}
U_+^\dagger U_- = U_-^\dagger U_+ = 0, \quad U_+^\dagger U_+ = U_-^\dagger U_- = 1, \\
\alpha U_\pm = \pm U_\pm, \quad \beta U_\pm = U_\mp.
\end{align}
\end{subequations}  
A perfectly valid representation would be:
\begin{align}
\gamma^0 &= \begin{pmatrix}
0 & 1 \\
1 & 0 
\end{pmatrix}\, , \quad \gamma^1 = \begin{pmatrix}
0 & -1 \\
1 & 0 
\end{pmatrix}\, ,  \quad\alpha = \begin{pmatrix}
1 & 0 \\
0 & -1 
\end{pmatrix}\, , \quad \beta = \gamma^0,  \\
U_+ &= \begin{pmatrix} 1 \\ 0 \end{pmatrix}\, ,\quad U_-= \begin{pmatrix} 0 \\ 1 \end{pmatrix}\, .
\end{align}
Our treatment will be independent of the representation and will be based on the properties defined in \eqref{eq:real-spin-basis}.

Complete sets of positive- and negative-frequency plane-wave solutions to Equation~\eqref{eq:2D-Dirac-eq} are \cite{Louko:2016ptn}: 
\begin{subequations}\label{eq:LinearInd-solns}
\begin{align}
\psi_{\pm, k}(t, x) &= \frac{1}{\sqrt{2\pi}} \frac{1}{\sqrt{2\omega_k}} u_\pm(k) e^{\mp i\omega_k t \pm ikx}, \\
u_\pm(k) &= \frac{\pm k_\mu \gamma^\mu + M}{\sqrt{2M(\omega_k + M)}} u_\pm(0), \quad u_\pm(0) = \sqrt{M}(U_+ \pm U_-),
\end{align}
\end{subequations}  
where $\omega_k = \sqrt{M^2 + k^2}$, and $u_\pm(k)$ are normalized such that:  
\begin{align}
u_\pm^\dagger(k) u_\pm(k) = 2\omega_k, \quad u_+^\dagger(k) u_- (k) = 0.
\end{align}  
Notably, 1+1D Dirac spinors lack spin degeneracy (unlike 3+1D), as the eigenstates $u_\pm(0)$ of $\gamma^0\equiv \beta$ are non-degenerate. 
The mode functions are normalised such that:
\begin{align}
(\psi_{s,k}, \psi_{s',k'}) : = \int_{-\infty}^{\infty} \psi^\dagger_{s,k}(t,x) \psi_{s',k'}(t,x)dx =\delta_{s,s'}\delta(k-k')\, ,\quad s,s' = \pm\, .
\end{align}

The quantized massive field is expanded in terms of creation/annihilation operators:  
\begin{align}\label{eq:External-quantized-spinor}
\psi(t, x) = \int_{-\infty}^\infty dk \left( b_{+, k} \psi_{+, k}(t, x) + d_{-, k}^\dagger \psi_{-, k}(t, x) \right),
\end{align}  
where $b_{+, k}$ and $d_{-, k}^\dagger$ are  the fermionic annihilation and creation operators for particles and antiparticles, respectively, which satisfy the canonical anticommutation relations: 
\begin{align}
\{b_{+, k}, b_{+, k'}^\dagger\} = \{d_{-, k}, d_{-, k'}^\dagger\} = \delta(k - k'), \quad \text{all other anticommutators} = 0.
\end{align}  
The Fock space is built on the Minkowski vacuum $|0_M\rangle$, defined by $b_{+, k}|0_M\rangle_{\psi} = d_{-, k}|0_M\rangle_{\psi} = 0$ for all $k \in \mathbb{R}$.

\subsection{Quantization of the Massless Dirac Field in the Cavity}  
\label{sec:Massless-DField}

The confined massless spinor $\chi(t, x)$ (clock field) satisfies boundary conditions at the cavity walls ($x = x_\pm$). We first consider MIT bag boundary conditions \cite{MIT-bag-original}, which enforce strict confinement:  
\begin{align}\label{eq:MIT-bag-bcs}
(i \mp \beta \alpha) \chi(t, x_\mp) = 0.
\end{align}  
These are analogous to Dirichlet conditions for scalar fields \cite{Lorek} and ensure no probability flux leaks through the walls.  

For generality, we also consider probabilistic boundary conditions, which relax the requirement of strict vanishing of the field at the cavity boundaries while enforcing the vanishing of  the probability current to vanish at each wall \cite{Friis:2011yd, Friis:2013eva}:  
\begin{align}\label{eq:prob-current-bcs}
\chi_{(1)}^\dagger \alpha \chi_{(2)} \bigg|_{x = x_\pm} = 0,
\end{align}  
where $\chi_{(1)}$ and $\chi_{(2)}$ are eigenfunctions of the massless Dirac Hamiltonian ($M = 0$) on either side of the boundary. Such boundary conditions belong to a family of self-adjoint extensions  of the Dirac operator constrained by boundary data, and are parametrized by a phase angle 
$\theta \in [0, 2\pi)$ and a spectral parameter $s \in [0, 1)$. The parameter
$\theta$ introduces a phase shift between incoming and reflected components of the field at the boundary, while $s$ determines the allowed mode spectrum and reflects the boundary’s admittance properties. These parameters shape the set of orthonormal eigenfunctions admitted by the boundary, which for 
$s \in (0, 1)$, take the form \cite{Friis:2011yd}:  
\begin{align}\label{eq:mode-funct-ProCurBcs}
\chi_n(t, x) = \frac{1}{\sqrt{2l}} \left( U_+ e^{i\tilde{\omega}_n(x - x_-)} + e^{i\theta} U_- e^{-i\tilde{\omega}_n(x - x_-)} \right) e^{-i\tilde{\omega}_n t},
\end{align}  
with corresponding mode frequencies shifted to $\tilde{\omega}_n = \frac{(n + s)\pi}{l}$ ($n \in \mathbb{Z}$). For $s = 1/2$ and $\theta = \pi/2$, the mode functions reduce to those of the standard MIT bag boundary conditions. The parameter $s$ is taken to be strictly greater than zero ($s \in (0, 1)$)
in this work to avoid the limiting case of $s=0$, which would admit a zero-mode with  solution $\tilde{\omega}_0 = 0$. Such a zero mode is excluded from quantization in the present framework as is done in prior work (e.g., Ref.~\cite{Friis:2011yd}). We therefore take the limit $s \to 0^+$, when necessary. This choice ensures that all quantized modes carry non-zero energy and avoids the treatment of such a zero-frequency mode in the context of confined, accelerated Dirac fields, as its quantization is nontrivial for accelerated motion and for our choice of boundary conditions, and is left for future investigation.  Importantly, since the coupling to the massive external field occurs via terms proportional to $\tilde{\omega}_1$, the zero mode — having $\tilde{\omega}_0 = 0$ — does not contribute to our decay rate calculation.

The quantized massless field in the cavity is a superposition of these eigenmodes:  
\begin{align}\label{eq:quantized-cavity-field}
\chi(t, x) = \sum_{n \geq 0} b_n \chi_n(t, x) + \sum_{n < 0} d_n^\dagger \chi_n(t, x),
\end{align}  
where the non-vanishing anticommutation relations of the fermionic creation and annihilation operators are assumed to satisfy the usual algebra: 
\begin{align}\label{eq:Cavity-creal-ann-ACR}
\{b_n, b_m^\dagger\} = \{d_n, d_m^\dagger\} = \delta_{nm}.
\end{align}  
The cavity Fock space is built on the vacuum $|0\rangle_\chi$, defined by $b_n |0\rangle_\chi = d_n |0\rangle_\chi = 0$ for all $n \in \mathbb{Z}$.

This model provides a framework for probing the Unruh effect through the decay of an excited state within the cavity. The fermionic nature of both the confined massless field $\chi$ and the external massive field $\psi$ introduces essential structure through their statistics and the boundary conditions imposed on $\chi$ at the cavity walls. The interaction, mediated by the Hermitian Hamiltonian $H_{\text{int}}$, enables energy exchange between the two fields. The central quantity of study is the decay rate of an excited state of $\chi$, which provides a direct signature of the modified vacuum fluctuations—a signature that manifests as a thermal Unruh bath when the entire cavity is accelerated.

%
\subsection{Inertial cavity decay rate}
\label{sec:intertial_decay}
We compute the decay rate of a cavity at rest in Minkowski spacetime following the timelike killing vector $\partial_t$.

We take the  cavity field to be initially in the one-particle state of the mode $\chi_n$ given in \eqref{eq:mode-funct-ProCurBcs} and the external field to be in the Minkowski vacuum. Working to first-order perturbation theory in the coupling strength $g$, the probability amplitude for the decay rate into the cavity vacuum state $|0\rangle_{\chi}$ and an arbitrary final state $|\beta\rangle_{\psi}$ of the external field is given by:
\begin{align}
\mathcal{A}_\downarrow = -i \int_0^t dt' {}_{\psi}\langle \beta |\,{}_{\chi}\langle 0| H_{int} |1 \rangle_{\chi} | 0_M\rangle_{\psi} \, .
\end{align}
Note the interaction Hamiltonian \eqref{eq:Hint_Coupling} is linear in both the cavity and external fields. Therefore, this first-order process can only create one particle in the external field and annihilate one in the cavity. This restricts the non-zero matrix elements to final states $|\beta\rangle_{\psi}$ that are single-particle states. Furthermore, energy conservation must be satisfied for that one-particle state. 

Crucially, the interaction term $\bar{\psi}\chi$ annihilates the cavity particle and create a particle in the external field. The conjugate term, $\bar{\chi}\psi$ would create an antiparticle in the cavity  and annihilate an antiparticle in the external field. Since the initial state of the external field is the vacuum (which contains no antiparticles), this second process is forbidden. Thus, only the $\bar{\psi}\chi$ term  contributes to the decay amplitude.
 
Collecting \eqref{eq:Hint_Coupling}, \eqref{eq:External-quantized-spinor}, \eqref{eq:quantized-cavity-field} 
and summing over all unobservable final states of the external field (i.e., over all such one-particle states), the total decay probability reads:
\begin{align}
\mathcal{P}_\downarrow = \sum_\psi |\mathcal{A}_\downarrow|^2= g^2 \int_{-\infty}^{\infty} dk \, |\Gamma_{k1}|^2 \, ,
\end{align}
where $\Gamma_{kn}$ is defined as:
\begin{align}
\Gamma_{kn} := \int_{0}^{t} dt' \int_{x_-}^{x_-+l}  dx \,  \bar{\psi}_{+,k}(t',x)\chi_n(t',x)\, ,
\end{align}
where $\bar{\psi}_{+,k} := \psi^\dagger_{+,k}\beta$. 

Collecting \eqref{eq:LinearInd-solns} and \eqref{eq:mode-funct-ProCurBcs}, we obtain the expression: 
\begin{subequations}\label{ProDecaIntegral}
\begin{align}
\mathcal{P}_\downarrow &= \frac{g^2}{4\pi l}\int_{-\infty}^\infty\, dk \frac{\sin^2\left( \left(\omega_k -\tilde{\omega}_1\right)\frac{t}{2}\right)}{(\omega_k- \tilde{\omega}_1)^2}\frac{1}{\omega_k(\omega_k + M)}\Bigg\{\kappa_+ \Bigg[ \frac{4\sin^2\left(\left(k-\tilde{\omega}_1\right)\frac{l}{2}\right)}{(k-\tilde{\omega}_1)^2}  \notag\\   &+ \frac{4\sin^2\left(\left(k+\tilde{\omega}_1\right)\frac{l}{2}\right)}{(k+\tilde{\omega}_1)^2} \Bigg]    + 2\kappa \Bigg[ \frac{4\sin^2\left(\left(k-\tilde{\omega}_1\right)\frac{l}{2}\right)}{(k-\tilde{\omega}_1)^2} - \frac{4\sin^2\left(\left(k+\tilde{\omega}_1\right)\frac{l}{2}\right)}{(k+\tilde{\omega}_1)^2} \Bigg] \notag\\   &+ \kappa_-\frac{4\left[\cos(\pi s)+\cos(kl) \right]\cos(\theta -\pi s)}{(k^2-\tilde{\omega}_1^2)}    \Bigg\} \, ,
\end{align}
where $\tilde{\omega}_1 =\frac{1+s}{l}\pi$, with  $s$ and $\theta$ as defined in Section \ref{sec:Massless-DField} and where $\kappa_{\pm}$ and $\kappa$ are defined as:
\begin{align}
\kappa_{\pm}: =  (\omega_k + M)^2\pm k^2 \, , \quad \kappa: = k(\omega_k +M)\, . 
\end{align}
\end{subequations}
The coupling coefficients $\kappa_{\pm}$ and $\kappa$ quantify the strength of the system’s coupling to the decay channel. Notably, the integrand is even in $k$.

\subsection{Short-Time and Long-Time Limits}

For $t \ll 1/|\omega_k - \tilde{\omega}_1|$, the temporal factor in the decay probability integral simplifies to $t^2/4$. In this regime, the integral \eqref{ProDecaIntegral} reduces to a short-time behavior dominated by quadratic time dependence:  
\begin{align}
\mathcal{P}_\downarrow^{\text{short}} \propto g^2 t^2 \cdot C(s, \theta),
\end{align} 
where $C(s, \theta)$ depends on the boundary conditions imposed on the cavity walls.

For long times, the $k$-integral is dominated by the resonance region $\omega_k \approx \tilde{\omega}_1$ (i.e., $k \approx \pm \tilde{k} = \pm \sqrt{\tilde{\omega}_1^2 - M^2}$), where the integrand peaks. The width of the resonance is $\Delta k \sim 1/l$, so the integral samples a range $\sim 1/l$ in $k$-space.
   Using the asymptotic identity for large $t$:  
\begin{align}
\frac{\sin^2\left( (\omega - \tilde{\omega}_1)t/2 \right)}{(\omega - \tilde{\omega}_1)^2} \sim \frac{\pi t}{2} \delta(\omega - \tilde{\omega}_1) \quad \text{as} \quad t \to \infty,
\end{align}  
the leading-order decay probability becomes:
\begin{subequations}\label{eq:Asymptotic-InertialDecayRate}
\begin{align}
\mathcal{P}_\downarrow^{\text{long}} &= \frac{g^2t}{lA\left(\tilde{\omega}_1 +M\right)}\Bigg\{\left(\tilde{\omega}_1 +M +A\right)^2\, \frac{\sin^2\left(\left(A-\tilde{\omega}_1\right)\frac{l}{2} \right)}{(A-\tilde{\omega}_1)^2}  \\  &+    \left(\tilde{\omega}_1 +M -A\right)^2 \, \frac{\sin^2\left(\left(A+\tilde{\omega}_1\right)\frac{l}{2} \right)}{(A+\tilde{\omega}_1)^2}  \notag \\ &-  2\frac{\left(\tilde{\omega}_1+M\right)}{M}\left[\cos(\pi s)+ \cos(lA)\right]\cos(\theta -\pi s) \Bigg \} \notag\, ,
\end{align}
where
\begin{align}
A:=\sqrt{\tilde{\omega}_1^2-M^2}\, \, ( M < \tilde{\omega}_1)  , \quad \tilde{\omega}_1 := \frac{(1 + s)\pi}{l}\, .
\end{align}
\end{subequations}

In the weak-mass limit, characterized by either ($\tilde{\omega}_1 \gg M$ or $Ml\ll 1$), the leading-order long-time behavior simplifies to:
\begin{align}\label{DecayStationaryLongReg}
\mathcal{P}_\downarrow^{\text{long}} \approx  g^2 t l \left[ 1 + \frac{M l \sin(\pi s) \cos(\theta - \pi s)}{(1 + s)^2\pi^2} \right]. 
\end{align} 
Thus, the long-time decay rate is: 
\begin{align}\label{DecayRateStationaryLongReg}
\Gamma_{\text{in}} =  g^2l \left[ 1 + \frac{M l  \sin(\pi s) \cos(\theta - \pi s)}{(1 + s)^2\pi^2} \right]. 
\end{align} 
Note that these conditions ($\tilde{\omega}_1 \gg M$ or $Ml\ll 1$) are physically equivalent: since $\tilde{\omega}_1 = (1+s)\pi/l$, the requirement $\tilde{\omega}_1 \gg M$ implies $Ml \ll (1+s)\pi$, which is equivalent to $Ml \ll 1$ up to an order-unity factor since $s\in(0,1)$. Both express, the regime where the Compton wavelength $1/M$ of the external field is much larger than the cavity size $l$, rendering the external field effectively massless relative to the cavity's geometric scale. Consequently, the leading behavior 
$\Gamma_{\text{in}}  \sim g^2  l$ is identical in both cases, with the mass M contributing only subleading corrections.
The bracket remaining positive for $s \in (0,1)$ and $\theta \in [0, 2\pi]$ under the condition $(1 + s)\pi \gg M l$. For the MIT bag boundary condition ($s = 1/2$, $\theta = \pi/2$), the decay rate simplifies to: 
\begin{align}
\Gamma^{\text{MIT}}_{\text{in}} =  g^2  l \left(1 + \frac{4 M l}{9\pi^2} \right)\, . 
\end{align}

\subsubsection{Limit:  $Ml\rightarrow 0$}
For $Ml \to 0$, the leading-order long-time decay rate simplifies to:
\begin{align}
 \Gamma_{\text{in}} = g^2 l\, .
\end{align}  
Crucially, Lorek et al.'s bosonic coupling $\lambda$ (dimension $[M^2]$) \cite{Lorek} yields a mass-independent decay rate $\Gamma_{\text{in}}^{\text{boson}} \sim \lambda^2 l^3$, while ours is $\Gamma_{\text{in}}^{\text{fermion}} \sim g^2 l$.  
Both results share this universal feature—mass independence of short-cavity decay—despite different coupling dimensions. The dimensions cancel in the decay rate, ensuring physical consistency and highlighting the universality of stationary short-cavity dynamics for inertial observers.

\section{Quantization of Dirac Field in Rindler Spacetime}
\label{sec:rindler_quantization}

To analyze the decay rate of a uniformly accelerated cavity, we first study the quantization of the massive Dirac field in the full Rindler spacetime and then specialize to the massless field confined within the accelerated cavity. The quantization of Dirac fields in Rindler wedges of Minkowski spacetime ($1+1$D here) has been extensively studied in the literature \cite{Bautista:1993cn,Candelas:1978gg,Jauregui:1991me,Hacian:1986hq,Takagi:1986kn,McMahon:2006ui,Soffel:1980kx, Boulware:1975pe}, with a focus on separating modes in the right (R: $x > |t|$) and left (L: $x < -|t|$) wedges. 

\subsection{Quantization of the Massive Dirac Field in Rindler Spacetime}
\label{sec:MassDiracQuantization-Rindler}
We adopt Rindler coordinates  $(\eta, \rho)$ to parametrize the right Rindler wedge R of Minkowski spacetime. These are defined in terms of Cartesian coordinates $(t,x)$ via:
\begin{align}\label{RindlerCoords}
t = \rho\sinh(a\eta) \, ,  \quad  \quad  x = \rho\cosh(a\eta) \, ,
\end{align}
where $0 < \rho < \infty$ and $-\infty < \eta< \infty$. Here, $a$ is a constant with
dimensions of inverse length, serving as a reference scale for proper acceleration. In these coordinates, the Minkowski metric takes the form $ds^2 = (a\rho)^2 d\eta^2 - d\rho^2$. An observer at fixed spatial coordinate $\rho =\rho_0$ follows a static worldline and experiences a proper acceleration given by $A = 1/\rho_0$. In particular, for the fiducial observer at $\rho_0 = 1/a$, the proper acceleration is $A =a$, and the Rindler time $\eta$ coincides with the observer’s proper time $\tau$, since $d\tau = (a\rho_0)d\eta =d\eta$. More generally, for an observer at arbitrary $\rho$, the proper time $d\tau$ relates to Rindler time via $d\tau = (a\rho) d\eta$.

In this geometry, the Dirac equation for the external massive spinor field can be written as \cite{Langlois:2005if,Crispino2008, Richter:2017noq}:  
\begin{align}
ia^{-1}\partial_{\eta} \psi(\eta, \rho) = \left(-i\alpha\left(\rho\partial_\rho + \frac{1}{2}\right) + M \rho\beta \right) \psi(\eta, \rho)\, ,
\end{align}
where $\alpha, \beta$ are spinor matrices defined in Section \ref{sec:MassiveDiracField}, and $M$ is still the mass of the external Dirac field.  

Working in the two-real components spinor basis \eqref{eq:real-spin-basis}, the normalized positive and negative frequency modes in the right Rindler wedge (R: $x > |t|$) are given by\footnote{These mode functions are equivalent to those found in Refs.\cite{Vanzella:2001ec,Richter:2017noq} under a specific spinor basis transformation; see end of subsection for details.}: 
\begin{subequations}\label{ModeFunct:RightWedge}
\begin{align}
\Psi_{+,\Omega}^R(\eta, \rho) &= \sqrt{\frac{M\cosh(\pi\Omega/a)}{a\pi^2}} \left( K_{i\frac{\Omega}{a}-\frac{1}{2}}(M\rho)U_{+} + i K_{i\frac{\Omega}{a}+\frac{1}{2}}(M\rho)U_{-} \right) e^{-i\Omega\eta}, \label{ModeFunct:RightWedge_1} \\
\Psi_{-,\Omega}^R (\eta, \rho) &= \sqrt{\frac{M\cosh(\pi\Omega/a)}{a\pi^2}} \left( K_{i\frac{\Omega}{a}-\frac{1}{2}}(M\rho)U_{-} + i K_{i\frac{\Omega}{a}+\frac{1}{2}}(M\rho)U_{+} \right) e^{i\Omega\eta},
\label{ModeFunct:RightWedge_2}
\end{align}
\end{subequations}
where $\Omega > 0$, $K_\nu(z)$ denotes the modified Bessel function of the second kind (decaying exponentially for large $z$, ensuring normalizability), and the subscript R labels right-Rindler wedge modes. We emphasize that negative frequency modes are obtained via charge conjugation of positive frequency modes. For a spinor $\psi$, charge conjugation is defined as $\psi^c = C\psi^*$, where $C$ is a matrix satisfying $C^\dagger C = \mathbb{I}$ and $C^\dagger \gamma^\mu C = -\gamma^{\mu*}$ \cite{Parker:2009uva}. In this $(1+1)$-dimensional framework, valid choices include $C = \gamma^1 = \beta\alpha$ or $C = -i\gamma^1 = -i\beta\alpha$ (with $\alpha, \beta$ as defined in Section \ref{sec:MassiveDiracField}); we use the latter to avoid spurious phase factors in the mode expansions. Modes are normalised such that: 
\begin{align}
(\Psi^R_{s,\Omega}, \Psi^R_{s',\Omega'}) : = \int_0^{\infty}d\rho \Psi^{R\dagger}_{s,\Omega}(\eta,\rho)\Psi^{R}_{s',\Omega'}(\eta,\rho)=\delta_{s,s'}\delta(\Omega -\Omega') \, , \quad s,s' =\pm  \, ,
\end{align}
where we used the properties of the spinors basis defined in \eqref{eq:real-spin-basis}, the identity $K_{\nu}(z)=K_{-\nu}(z)$ and the following important Modified Bessel identity derived in Appendix A of \cite{Falcone:2023tqw}:
\begin{align}
\int_0^{\infty}d\rho K_{i\frac{\Omega}{a}+\frac{1}{2}}(M\rho) K_{i\frac{\Omega'}{a}-\frac{1}{2}}(M\rho)+ K_{i\frac{\Omega}{a}-\frac{1}{2}}(M\rho) K_{i\frac{\Omega'}{a}+\frac{1}{2}}(M\rho)= \frac{a}{M}\frac{\pi^2 \delta(\Omega -\Omega')}{\cosh(\pi\Omega/a)}\, .
\label{eq:MBesselIdentity}
\end{align}

In the left Rindler wedge (L: $x < -|t|$), we extend the right-Rindler modes by setting $\Psi_{\pm,\Omega}^R = 0$ in L. To describe physics in L, we adopt Rindler coordinates $(t, x) = (-\rho\sinh(a\eta), -\rho\cosh(a\eta))$ (with $\rho > 0$, $\eta \in \mathbb{R}$), preserving consistency with R’s labeling. 

The left-Rindler modes, nonvanishing in L and vanishing in R, are defined with a sign flip in the frequency dependence (due to $\partial_\eta$ being past-directed in L):  
\begin{subequations}\label{ModeFunct:LeftWedge}
\begin{align}
\Psi_{+,\Omega}^L(\eta, \rho) &= \sqrt{\frac{M\cosh(\pi\Omega/a)}{a\pi^2}} \left( K_{i\frac{\Omega}{a}-\frac{1}{2}}(M\rho)U_{+} + i K_{i\frac{\Omega}{a}+\frac{1}{2}}(M\rho)U_{-} \right) e^{i\Omega\eta}, \\
\Psi_{-,\Omega}^L (\eta, \rho) &= \sqrt{\frac{M\cosh(\pi\Omega/a)}{a\pi^2}} \left( K_{i\frac{\Omega}{a}-\frac{1}{2}}(M\rho)U_{-} + i K_{i\frac{\Omega}{a}+\frac{1}{2}}(M\rho)U_{+} \right) e^{-i\Omega\eta}.
\end{align}
\end{subequations}
These left modes are positive/negative frequency with respect to $-\partial_\eta$ (the Rindler time coordinate in L) and satisfy $(\Psi^L_{s,\Omega}, \Psi^L_{s',\Omega'}) = \delta_{s,s'} \delta(\Omega - \Omega')$.  

The combined set of left and right Rindler modes form a complete set in $R\cup L$. Thus, the quantized massive Dirac field  in this region can be written as the expansion: 
\begin{align}\label{ExternalField-Rindler}
\Psi = \int_0^\infty d\Omega\, \left( b_{+,\Omega}^R \Psi_{+,\Omega}^R  + d_{-,\Omega}^{R\dagger} \Psi_{-,\Omega}^R + b_{+,\Omega}^L \Psi_{+,\Omega}^L + d_{-,\Omega}^{L\dagger} \Psi_{-,\Omega}^L \right)\equiv \Psi_R + \Psi_L,
\end{align}  
where $b_{+,\Omega}^P, b_{+,\Omega}^{P\dagger}$ ($d_{-,\Omega}^P, d_{-,\Omega}^{P\dagger}$) denote annihilation/creation operators for Rindler particles/antiparticles in wedge P = R, L. The nonvanishing anti-commutation relations are:  
\begin{align}
\lbrace b_{+,\Omega}^P, b_{+,\Omega'}^{P\dagger} \rbrace = \lbrace d_{-,\Omega}^P, d_{-,\Omega'}^{P\dagger} \rbrace = \delta(\Omega - \Omega'), \quad \text{P = L, R}.
\end{align}  
The Fulling-Rindler (FR) vacuum $|0_{\text{FR}}\rangle \equiv|0_{\text{R}},0_{\text{L}}\rangle  $ defined as the state annihilated by all Rindler operators ($b_{+,\Omega}^P |0_{\text{FR}}\rangle = d_{-,\Omega}^P |0_{\text{FR}}\rangle = 0$ for P = R, L), is empty for Rindler observers in $\text{R}\cup \text{L}$. 

Under the assumption that the entanglement in the Minkowski vacuum couples only Rindler modes of the same frequency $\Omega$ across the left and right wedges  \cite{Takagi:1986kn, Birrell:1982ix, Alsing:2006cj}, the Minkowski vacuum $|0_M\rangle$ is related to $|0_{\text{FR}}\rangle$ by a unitary two-mode squeezing operator:  
\begin{align}\label{SqueezingOperator}  
S = \exp\left( \int_0^\infty d\Omega \, r_\Omega \left( b_{+,\Omega}^{R\dagger} b_{+,\Omega}^{L\dagger} - b_{+,\Omega}^R b_{+,\Omega}^L \right) \right),  
\end{align}  
where $r_\Omega = \arctan(e^{-\pi\Omega/a})$. 
The action of $S$ on the FR vacuum produces two-mode squeezed states:  
\begin{align}  
|0_M\rangle = S |0_{\text{FR}}\rangle = \bigotimes_\Omega |\psi_\Omega\rangle, \quad |\psi_\Omega\rangle = \cos r_\Omega |0_R, 0_L\rangle_\Omega + \sin r_\Omega |1_R, 1_L\rangle_\Omega,  
\end{align}  
where $|n_R, n_L\rangle_\Omega = (b_{+,\Omega}^{R\dagger})^{n_R} (b_{+,\Omega}^{L\dagger})^{n_L} |0_R, 0_L\rangle_\Omega$ ($n_R, n_L \in \{0,1\}$). Notably, the key difference from the scalar case is that the fermionic state is a product over frequencies of two-mode states, each of which is a superposition of the vacuum and the doubly occupied state (one particle in R and one in L). There are no higher occupation numbers because of the Pauli exclusion principle.
The transformed annihilation/creation operators satisfy:  
\begin{subequations}\label{eq:TransformedRindlerOperators}  
\begin{align}  
S^\dagger b_{+,\Omega}^R S &= \cos r_\Omega \, b_{+,\Omega}^R - \sin r_\Omega \, b_{+,\Omega}^{L\dagger}, \\  
S^\dagger b_{+,\Omega}^L S &= \cos r_\Omega \, b_{+,\Omega}^L + \sin r_\Omega \, b_{+,\Omega}^{R\dagger}.  
\end{align}  
\end{subequations}

In Refs.\cite{Vanzella:2001ec,Richter:2017noq}, the positive frequency mode  function is of the form: 
$$\Psi_{+,\Omega}^R(\eta, \rho)= N_\Omega \begin{pmatrix} K_{i\frac{\Omega}{a}+\frac{1}{2}}(M\rho) + iK_{i\frac{\Omega}{a}-\frac{1}{2}}(M\rho) \\ -K_{i\frac{\Omega}{a}+\frac{1}{2}}(M\rho) + iK_{i\frac{\Omega}{a}-\frac{1}{2}}(M\rho) \end{pmatrix} e^{-i\Omega\eta}\, , \quad N_\Omega=\sqrt{\frac{M\cosh(\pi\Omega/a)}{2a\pi^2}}\, .$$
This is equivalent to our solution \eqref{ModeFunct:RightWedge_1} under the (complex) spinor basis transformation $U_+\leftrightarrow\tilde{U}_+=\begin{pmatrix} i \\ i \end{pmatrix}$ and $U_-\leftrightarrow \tilde{U}_-=\begin{pmatrix} -i \\ i \end{pmatrix}$. Both solutions satisfy Dirac equation;  but their spinor basis is orthogonal but non-normalized (i.e., $\tilde{U}_+^\dagger \tilde{U}_+=\tilde{U}_-^\dagger \tilde{U}_- = 2$, and $\tilde{U}_+^\dagger \tilde{U}_-=\tilde{U}_-^\dagger \tilde{U}_+ = 0$), resulting in an additional factor of $1/\sqrt{2}$ in their mode functions normalization factor $N_\Omega$. The two choices are physically equivalent in the sense that physical observables depend on $N_\Omega^2 \tilde{U}_s^\dagger \tilde{U}_{s'}$ for $s,s'=\pm$, which is the same in both bases.

\subsection{Massless Dirac Field in an Accelerated Cavity}
\label{sec:AcceleratedCavity-Field}

We now specialize the general quantization scheme for the right Rindler wedge (R) to the case of a massless Dirac field confined within a cavity undergoing uniform acceleration.

Consider a cavity confined to the right Rindler wedge R with walls dragged along the boost Killing vector $K = x\partial_t + t\partial_x$. The worldlines of the cavity walls are $x = \sqrt{x_-^2 + t^2}$ (proper acceleration $a_- = 1/x_-$) and $x = \sqrt{x_+^2 + t^2}$ (proper acceleration $a_+ = 1/x_+$), overlapping with the inertial cavity at $t = 0$. In R, the cavity walls are, respectively, at $\rho_-=\rho_0$ and $\rho_+=\rho_0 +l$ (with $\rho_0 =x_-$ at t=0).

Using boundary conditions analogous to Eq.~\eqref{eq:prob-current-bcs} with $x_{\mp}\rightarrow \rho_{\mp}$, the normalized mode solutions of the massless Dirac equation within the accelerated cavity confined to (R) are\cite{Friis:2011yd}:  
\begin{subequations}\label{CavityFieldRRW}  
\begin{align}  
\chi_n^R(\eta, \rho) &= \frac{e^{-i\Omega_n \eta} \left( \left( \frac{\rho}{\rho_-} \right)^{\frac{i\Omega_n}{a}} U_+ + e^{i\theta} \left( \frac{\rho}{\rho_-} \right)^{-i\frac{\Omega_n}{a}} U_- \right)}{\sqrt{2\rho \ln(\rho_+/\rho_-)}}, \\  
\Omega_n &:= \frac{a(n + s)\pi}{\ln(\rho_+/\rho_-)}, \quad n \in \mathbb{Z},  
\end{align}  
\end{subequations}  
where $\theta$ and $s$ are defined as in Section~\ref{sec:Massless-DField}. The normalization factor $1/\sqrt{2\rho \ln(\rho_+/\rho_-)}$ ensures orthonormality over the cavity volume, accounting for the logarithmic dependence of the proper length on $\rho$.

The quantized massless Dirac field within the accelerated cavity is:  
\begin{align}\label{quantized-cavity-field_Rindler}  
\chi(\eta, \rho) = \sum_{n \geq 0} B_n \chi_n^R(\eta, \rho) + \sum_{n < 0} D_n^\dagger \chi_n^R(\eta, \rho),  
\end{align}  
with anticommutation relations:  
\begin{align}\label{Cavity-creal-ann-ACR_Rindler}  
\{ B_n, B_m^\dagger \} = \{ D_n, D_m^\dagger \} = \delta_{nm}.  
\end{align} 
The vacuum state $|0\rangle_{\chi^R}$—empty for accelerated observers—is defined by $B_n |0\rangle_{\chi^R} = D_n |0\rangle_{\chi^R} = 0$ for all $n \in \mathbb{Z}$. This vacuum differs from the FR vacuum due to the cavity boundaries (Eqs.~\eqref{eq:MIT-bag-bcs}-\eqref{eq:prob-current-bcs}), which break conformal invariance and induce discrete mode structure.

\subsection{Choice of Proper Time and the Small Cavity Approximation}
\label{sec:ProperTimeChoice}
The quantization procedure in Sec.~\ref{sec:MassDiracQuantization-Rindler} and Sec.~\ref{sec:AcceleratedCavity-Field} employ Rindler time $\eta$ which is a coordinate time. To connect with physical observables such as decay rates, we must express dynamics in terms of the proper time $\tau$
of a specific observer. Two natural approaches present themselves for this analysis.

In the first approach, we consider a fiducial observer at the left wall, setting $\rho_- = \rho_0 = 1/a$ such that this observer has proper acceleration $a$. For this observer, the proper time is given by $d\tau = a \rho_- d\eta =d\eta$ (since $a\rho_-=1)$, making Rindler time coincide with their proper time. This identification is particularly useful under the condition $l \ll \rho_- = 1/a$, where the variation in proper time rate across the cavity becomes negligible. Under this condition, the proper frequency of the internal massless cavity field, measured by this observer, becomes $\Omega_n  \approx \pi(n+s)/l=\tilde{\omega}_n $, causing the acceleration parameter to cancel out and rendering the spectrum identical to that of an inertial cavity of proper length $l$. This first approach provides a direct link to an observer (or detector) fixed to the cavity wall.
If we relax the geometric condition $l\ll 1/a$, and only rely on the external field condition ($Ml \ll 1$), we enter a regime where non-trivial dependence on $a$ emerges.

Alternatively, one may consider an observer at the center of the cavity, located at $\rho_c = (\rho_- +\rho_+)/2$, with proper time $d\tau = a\rho_c d\eta$.
This approach is more broadly applicable as it does not require the cavity to be small relative to $\rho_-$ or $1/a$. From the mode solutions \eqref{CavityFieldRRW}, the proper frequency (as measured at the center) is $\Omega_n =(n+s)\pi/(\rho_c \ln(\rho_+/\rho_-))$, where $\rho_{\pm} = \rho_c \pm l/2$.  For a cavity that is small relative to its position, $l \ll \rho_c$, the proper frequency measured by this observer is also $\Omega_n \approx \pi(n+s)/l$. Again, the internal energy level structure decouples from the acceleration scale.

The decoupling of the acceleration scale from the internal energy level structure is a universal feature of a small accelerated cavity  $al \ll 1$, regardless of which observer's proper time we use. The physical reason is that for a very small cavity, the tidal forces (the variation of proper acceleration across it) become negligible. The entire cavity behaves almost like a point-like object undergoing rigid acceleration. Its internal structural properties, such as the energy gaps $\Delta E =\hbar \pi/l $ between successive modes, become independent of its overall motion. The decay probability, which depends on these energy gaps, will therefore be identical. The remaining effect of acceleration is contained in the definition of the vacuum state $|0\rangle_{\chi^R}$, which differs from the Minkowski vacuum and exhibits thermal properties due to the Unruh effect.
This regime provides a natural testbed for testing Unruh effect.

For the decay analysis in the following section, we first adopt a general approach, using an arbitrary  proper time, later for the asymptotics of the decay rate we specialize to a fiducial observer at the left wall.

\section{Uniformly Accelerated Cavity: Decay Rate Analysis}
\label{sec:accelerated_decay}

In this section, we analyze the decay probability of a fermionic particle confined to an accelerated cavity mode, leveraging the causal structure of Rindler spacetime and the constraints of quantum field theory to derive the expression for the decay rate. Our analysis emphasizes the role of spatial disjointness between Rindler wedges, particle number conservation, and the structure of the interaction Hamiltonian, ensuring consistency with the Unruh effect in curved spacetime. We note that the term `particle number conservation' used here refers to the conservation of total excitations in the interaction process between the cavity and external fields, as described in the accelerated frame's Fock space. The decay is stimulated by the Unruh thermal bath perceived by the accelerated cavity, which drives transitions from cavity excitations to excitations in the external field. Our analysis focuses on such stimulated transitions between specific states, not on vacuum particle production, and the particle number is defined with respect to the Rindler mode expansion appropriate for the accelerated cavity. This is distinct from the Unruh effect, where an accelerated observer detects particles in the Minkowski vacuum due to the different definition of particles in Rindler versus inertial quantization.

The system is initialized with one fermion in the fundamental mode $(n=1)$ of a cavity localized to the right Rindler wedge (R), denoted $|1\rangle_{\chi^R}$, and the external Dirac field in the Minkowski vacuum $ |0_M\rangle_{\psi}$.
Using the squeezing operator $S$, which maps the Fulling-Rindler vacuum to Minkowski vacuum, the initial state reads $|i\rangle = |1\rangle_{\chi^R} \otimes |0_M\rangle_{\psi} = |1\rangle_{\chi^R} \otimes S |0_{FR}\rangle$. The final state is defined as an empty cavity ($|0\rangle_{\chi^R}$) and the external field in a Minkowski coherent state $|\beta_\Omega\rangle_M:=S|\beta_\Omega\rangle_R$, where $|\beta_\Omega\rangle_R$ is a Rindler coherent state of the external Dirac field in R with amplitude $\beta_\Omega$ for mode $\Omega$. This final state is $
|f\rangle = |0\rangle_{\chi^R} \otimes S|\beta_\Omega\rangle_R.
$  
This ensures consistency with the interaction Hamiltonian $H_{\text{int}}$, which acts in Minkowski spacetime. This mapping is critical for maintaining coherence between the accelerated frame of the cavity and the inertial frame of the external field.

Working to leading order in perturbation theory, the transition amplitude $\mathcal{A}_\downarrow^{R}$ describing the decay of the initial cavity state to the final external field state is given by:
\begin{align}
\mathcal{A}_\downarrow^{R} = -i \int_0^\tau \frac{d\tau'}{a\rho_0} {}_R\langle \beta_\Omega |\,{}_{\chi^R}\langle 0| S^\dagger H_{int}(\tau) S|1 \rangle_{\chi^R}| 0_{\text{FR}}\rangle \, ,
\end{align}
where $\tau$ is the total proper interaction time for an observer at fixed $\rho =\rho_0$, while the measure $\frac{d\tau'}{a\rho_0}$ properly accounts for the conversion from Rindler time $\eta$ to the proper time $\tau$ of the observer at fixed $\rho_0$. For an observer at the center of the cavity the proper time $\tau_c = a\rho_c \eta$ while for a fiducial left-wall observer with $\rho =\rho_- =1/a$, the proper time is $\tau_- = a\rho_-\eta =\eta$.

The causal separation of the right (R: $x > |t|$) and left (L: $x < -|t|$) Rindler wedges ensures cross-wedge interactions vanish, simplifying the decay amplitude to an intra-wedge term only. This follows because modes in R (e.g., the cavity mode $\chi^R_1$) and L (all external modes) have disjoint supports: $\text{supp}(\Psi^R_{\pm\Omega}) \cap \text{supp}(\Psi^L_{\pm\Omega}) = \emptyset$ and $\text{supp}(\chi^R_1) \cap \text{supp}(\Psi^L_{\pm\Omega}) = \emptyset$. Modes are explicitly defined to vanish in the opposite wedge, enforcing no overlap.  

Furthermore, the local interaction $H_{\text{int}} \propto \bar{\chi}\psi + \bar{\psi}\chi$ conserves the total number of fermions (cavity + external). This imposes a critical constraint: the transition amplitude $\mathcal{A}_\downarrow^{R}$ can only involve states with identical fermion numbers in R.  
Collecting the mode expansions \eqref{ExternalField-Rindler}, \eqref{quantized-cavity-field_Rindler}, we obtain the explicit form: 
\begin{align}
\mathcal{A}_{\downarrow}^{R}= -ig \int_0^\tau \frac{d\tau'}{a\rho_0}  \int_{\rho_-}^{\rho_+} d\rho \int_0^\infty d\Omega \bar{\Psi}^R_{+,\Omega}\chi^R_1 \cos r_\Omega \,{}_R\langle \beta_\Omega| b^{R\dagger}_{+,\Omega}|0_{\text{FR}}\rangle\, . 
\end{align}

The amplitude contains several critical components. First, the mode functions $\chi^R_1(\tau', \rho)$ and $\bar{\Psi}^R_{+,\Omega}(\tau', \rho)$ describe the cavity and external field modes, respectively. The cavity mode $\chi^R_1$ is normalized and localized exclusively to the cavity volume $\rho_- \leq \rho \leq \rho_+$, ensuring it does not overlap with the left wedge. The external field mode $\bar{\Psi}^R_{+,\Omega}$ is the adjoint of the Dirac field mode in R, responsible for creating a fermion in mode $\Omega$.

Second, the phase factor $\cos r_\Omega$ arises from the Bogoliubov transformation \eqref{eq:TransformedRindlerOperators}, which diagonalizes the Hamiltonian in the accelerated frame. For fermions, this transformation introduces a thermal distribution, with $r_\Omega$ encoding the Unruh effect. This reflects the thermal nature of the Minkowski vacuum as perceived by the accelerated observer.

Third, the coherent state overlap ${}_R\langle \beta_\Omega | b^{R\dagger}_{+,\Omega} | 0_{\text{FR}} \rangle$ quantifies the overlap between the Rindler coherent state $|\beta_\Omega\rangle_R$ and the 1-particle state $|1_\Omega\rangle_R$ created by the Rindler creation operator $b^{R\dagger}_{+,\Omega}$ acting on the Fulling-Rindler vacuum $|0_{\text{FR}}\rangle$. For a coherent state, this overlap is non-zero only for the 1-particle component, given by $\beta_\Omega e^{-|\beta_\Omega|^2/2}$, where $\beta_\Omega e^{-|\beta_\Omega|^2/2}$ is the overlap between $|\beta_\Omega\rangle_R$ and $|1_\Omega\rangle_R$.

\subsection{Decay Probability in Right Rindler Wedge}

The decay probability $\mathcal{P}_\downarrow^{R}$ is obtained by summing the squared modulus of the transition amplitude over all final Rindler coherent states $|\beta_\Omega\rangle_R$. Particle number conservation (initial 1-particle cavity state) enforces that only the 1-particle component of $|\beta_\Omega\rangle_R$ contributes, as higher-number states ($n \geq 2$) cannot be created from the vacuum.

Summing over $\beta_\Omega$ (equivalent to integrating over the external field’s density matrix) and using the normalization of coherent states, the decay probability becomes:
\begin{subequations}\label{DecayProb-Rindler-1}
\begin{align}
\mathcal{P}_\downarrow^{R} = g^2 \int_0^\infty d\Omega \, \left| \Gamma_1^{R}(\Omega) \right|^2 \frac{e^{2\pi\Omega/a}}{e^{2\pi\Omega/a} + 1}\, ,
\end{align}
where $\Gamma_n^{R}(\Omega)$ is the intra-wedge overlap integral defined as:
\begin{align}
\Gamma_n^{R}(\Omega) := \int_{0}^{\tau} \frac{d\tau'}{a\rho_0}  \int_{\rho_-}^{\rho_+}  d\rho \,  \bar{\psi}_{+,\Omega}^R(\tau',\rho)\chi^R_n(\tau',\rho)\, ,
\end{align}
\end{subequations}
and where $\bar{\psi}_{+,\Omega}^R(\tau',\rho) :=\psi_{+,\Omega}^{R\dagger}(\tau',\rho)\beta$. The thermal factor $\frac{e^{2\pi\Omega/a}}{e^{2\pi\Omega/a} + 1} = \frac{1}{1+e^{-2\pi\Omega/a}}$ represents Fermi-Dirac stimulation from the Unruh thermal bath at the temperature $T=a/(2\pi)$.

Collecting the mode functions Eqs. \eqref{ModeFunct:RightWedge_1} and \eqref{CavityFieldRRW}, we obtain:
\begin{subequations}\label{DecayProb-Rindler-1}
\begin{align}
\mathcal{P}_\downarrow^{R} &= \frac{g^2 M}{a \pi^2\ln\left(\frac{\rho_- + l}{\rho_-}\right)}\left(\frac{1}{a\rho_0}\right)^2 \int_0^\infty d\Omega \frac{e^{\pi\Omega/a}\sin^2\left( \left(\Omega -\Omega_1\right)\frac{\tau}{2a\rho_0}\right)}{\left(\left(\Omega -\Omega_1\right)\frac{1}{a\rho_0}\right)^2}\left|f(\Omega) \right|^2 \, , 
\end{align}
where
\begin{align}\label{eq:Mode Overlap Rindler}
f(\Omega) := \int_{\rho_-}^{\rho_- + l}  \frac{d\rho}{\sqrt{\rho}} \left(e^{i\theta}\left(\frac{\rho}{\rho_-}\right)^{-i\frac{\Omega_1}{a}}K_{i\frac{\Omega}{a} +\frac{1}{2}}(M\rho)-i\left(\frac{\rho}{\rho_-}\right)^{i\frac{\Omega_1}{a}}K_{i\frac{\Omega}{a} -\frac{1}{2}}(M\rho)  \right)\, ,
\end{align}
and where 
\begin{align}
\Omega_1 := \frac{a(1+s)\pi}{\ln\left(\frac{\rho_- + l}{\rho_-}\right)}, \quad s\in(0,1), \quad \theta\in[0,2\pi)\, .
\end{align}
\end{subequations}

\subsection{Long-Time Resonance Dominance ($\tau \gg 1/\Omega_1$)}  
For long measurement times ($\tau \gg 1/\Omega_1$), the integrand of \eqref{DecayProb-Rindler-1} is dominated by the resonance condition $\Omega \approx \Omega_1$, where the sine-squared term $\sin^2\left( (\Omega - \Omega_1)\tau/2 \right)$ peaks sharply. Using the asymptotic identity for large $\tau$:  
\begin{align}
\frac{\sin^2\left( \left(\Omega -\Omega_1\right)\frac{\tau}{2a\rho_0}\right)}{\left(\left(\Omega -\Omega_1\right)\frac{1}{a\rho_0}\right)^2}\sim a\rho_0\frac{\pi \tau}{2} \delta(\Omega - \Omega_1) \quad \text{as} \quad \tau \to \infty,
\end{align}  
the leading-order decay probability becomes:  
\begin{equation}
\mathcal{P}_\downarrow^{R\text{long}} = \frac{g^2  M e^{\pi \Omega_1/a}}{2a\pi \ln\left(\frac{\rho_- + l}{\rho_-}\right)} \frac{\tau}{a\rho_0} \left|f(\Omega_1) \right|^2\, , \label{eq:long_time_limit}
\end{equation}  
where $\tau$ is the total proper time of the observer chosen to define the calculation, and $\Delta \eta =  \frac{\tau}{a\rho_0}$ is the corresponding total Rindler time interval, which is the same for all observers.

Notably, for a fiducial observers at the left wall of the uniformly accelerated cavity ($\rho_0 =\rho_- =1/a$), their proper time is given by $d\tau_{\text{left}} = a\rho_- d\eta = d\eta$. Therefore, the total Rindler time $\Delta \eta$ corresponds to a total proper time $\tau_{\text{left}} =\Delta \eta$.
Their measured decay rate is:
\begin{align}\label{eq:DecayRate-LeftWallObs_R}
\Gamma_{\text{left}} =  \frac{\mathcal{P}^{\text{Rlong}}_\downarrow}{ \tau_{\text{left}}} = \frac{\mathcal{P}^{\text{Rlong}}_\downarrow}{\Delta \eta} =\frac{g^2  M e^{\pi \Omega_1/a}}{2a\pi \ln\left(1+al\right)} \left|f(\Omega_1) \right|^2\, . 
\end{align}

For observers at the center of the cavity ($\rho_0 =\rho_c, \rho_{\pm} = \rho_c\pm l/2$), their proper time is $d\tau_c= a\rho_c d\eta$. Therefore, the same total Rindler time interval $\Delta \eta$ corresponds to a total proper time $\tau_c = a\rho_c \Delta\eta$. Their measured decay rate is:
\begin{align}
\Gamma_c = \frac{\mathcal{P}^{\text{Rlong}}_\downarrow}{\Delta \tau_{\text{center}}} = \frac{\mathcal{P}^{\text{Rlong}}_\downarrow}{a \rho_c \Delta \eta} = \frac{g^2  M e^{\pi \Omega_1/a}}{2a^2 \rho_c\pi \ln\left(\frac{\rho_c + l/2}{\rho_c-l/2}\right)} \left|f(\Omega_1) \right|^2\, .
\end{align}
The factor $1/(a\rho_c)$ in $\Gamma_{\text{center}}$ arises directly from time dilation, since  the central observer's clock runs faster relative to Rindler time.

The two rates $\Gamma_{\text{left}}$ and $\Gamma_{\text{center}}$ differ because they are defined relative to different proper times. This is physically meaningful: the central observer's clock runs at a different rate compared to the left-wall observer's clock due to their different positions in the accelerated frame.  For the central observer, total probability $\mathcal{P}^{\text{Rlong}}_\downarrow$
unfold over a longer period of their own proper time, so the rate $(d\mathcal{P}^{\text{Rlong}}_\downarrow/d\tau)$ is smaller. 
The decay probability $\mathcal{P}^{\text{Rlong}}_\downarrow$ itself is a Lorentz scalar, the same for both observers, as it is calculated over the same physical process. However, the rate $\Gamma = d\mathcal{P}^{\text{Rlong}}_\downarrow/d\tau $ depends on whose proper time $\tau$ you use to measure it. 

For the remainder of this work, we will use the proper time of the left-wall observer ($\rho_0 =\rho_-=1/a$), which yield the decay rate $\Gamma_{\text{left}}$.

\section{Comprehensive Analysis of Decay Probability Limits}
\label{sec:results}
The decay probability $\mathcal{P}_\downarrow^{R\text{long}}$ in the long-time resonance limit ($\tau \gg 1/\Omega_1 $), for the uniformly accelerated cavity, depends on geometric parameters (cavity length $l$, position $\rho_-$), acceleration $a$, mass scale $M$, and measurement time $\tau$. Below, we analyze key limits governing its behavior.

\subsection{Decay Rate in the Short-Cavity Limit ($al \ll 1$)}
\label{sec:ShortCavity al<<1}
In this approximation, we model an experiment where a specific accelerated detector is attached to the left wall of our uniformly accelerated cavity. This detector prepares, manipulates, and measures the quantum field within its local cavity. The time parameter $\tau$ in the decay probability is the time read by this detector's clock. We consider the regime $al \ll 1$, which we established in Sec.~\ref{sec:ProperTimeChoice} as the condition where tidal forces are negligible. As discussed therein, this hierarchy of scales ($l \ll 1/a$) ensures the variation in proper time and acceleration across the cavity is negligible, allowing the entire cavity to be treated as a rigid object undergoing uniform acceleration. Consequently, the external field modes vary slowly across the cavity's finite extent, which will permit a significant simplification of the overlap integrals.

Under the condition $l \ll \rho_- =1/a$, the integration domain $[\rho_-, \rho_- +l]$ of the overlap function \eqref{eq:Mode Overlap Rindler}  is small. 
Although the modified Bessel functions  $K_{i\beta \pm 1/2}(M\rho)$ (where $\beta:= \Omega_1/a$) are highly oscillatory functions of $\rho$ for large $\beta$, their variation over the interval $l$ is negligible. Therefore, we approximate them by their value at the lower limit $\rho_0=\rho_- =1/a$. In this regime, the leading order of the overlap function \eqref{eq:Mode Overlap Rindler} is:
\begin{align}\label{ModeOverlap:shortCav-Rindler}
f(\Omega_1) \approx l\sqrt{a}\left[e^{i\theta}K_{i\beta +\frac{1}{2}}(M/a) -i K_{i\beta -\frac{1}{2}}(M/a)\right]\, .
\end{align} 

We now assume $M/a\ll 1 $ (the weak-mass regime). The short-cavity condition $al \ll 1$ implies $\beta = \frac{\Omega_1}{a} = \frac{(1+s)\pi}{al} \gg 1$. 
For fixed $\nu$ and $z\rightarrow 0$,  the modified Bessel function $K_{\nu}(z)$ has  the asymptotic expansion \cite{DLMF}: $K_\nu(z)\sim \frac{1}{2}\Gamma(-\nu) \left(\frac{z}{2}\right)^\nu + \frac{1}{2}\Gamma(\nu) \left(\frac{z}{2}\right)^{-\nu}$.  Using the asymptotic forms $\Gamma(\pm i\beta + 1/2)\sim\sqrt{2\pi}e^{-\pi\beta/2}e^{\pm i(\beta\ln\beta -\beta}$, we obtain the dominant behavior of the modulus squared:
\begin{align}\label{eq:OverlapMagnitude}
|f(\Omega_1)|^2 \sim \frac{2\pi l^2 a^2}{M}e^{-\pi\beta}\left[1 - \sin\left(\theta + 2\beta \ln\left(\frac{2 a\beta}{M} \right) -2\beta \right) \right]\, .
\end{align}
Combining this with the long-time probability \eqref{eq:long_time_limit} and using the  approximation $\ln (\rho_+/\rho_-)\approx al$ valid for $al \ll 1$, the decay probability becomes:
\begin{align}
\mathcal{P}^{\text{Rlong}}_\downarrow \sim  g^2 l\tau_{\text{left}} \left[1 - \sin\left(\theta + 2\beta \ln\left(\frac{2 \Omega_1}{M} \right) -2\beta \right) \right]\, .
\end{align}
The oscillating term  $\sin\left(\theta + 2\beta \ln\left(\frac{2 \Omega_1}{M} \right) -2\beta \right)$ 
is bounded between $-1$ and $1$. For $\beta \gg 1$, the phase $\phi:=\theta + 2\beta \ln\left(\frac{2 \Omega_1}{M} \right) -2\beta$ varies extremely rapidly with $\beta$ (and thus with the acceleration $a$).
Its derivative $d\phi/d\beta = 2 \ln\left(\frac{2 \Omega_1}{M} \right) -2$ is large. Consequently, any infinitesimal experimental uncertainty $\delta a$ in the acceleration  will randomize the phase modulo $2\pi$, leading to $\langle \sin\phi \rangle = 0$ upon averaging. Therefore, the averaged decay rate is:
\begin{equation}
\Gamma_{\text{left}} \sim  g^2 l = \Gamma_{\text{in}}\, .
\end{equation}
This result is exact in the idealized mathematical limit of a vanishingly small cavity ($al\rightarrow 0$). In this limit, the leading-order asymptotic expression converges exactly to $\Gamma_{\text{left}} =  g^2 l = \Gamma_{\text{in}}$, and all higher corrections vanish.

This result makes physical sense because: for $al \ll 1$, the cavity mode frequency $\Omega_1 \sim \pi/l$ is much larger than the acceleration scale $a$ (since $\Omega_1/a \sim \pi/(al) \gg 1$). The Unruh thermal bath at temperature $T= a/(2\pi)$ has an exponentially small population at frequency $\Omega_1:$ $n(\Omega_1)\sim e^{-2\pi\Omega_1/a}$. Thus, stimulated emission is negligible, and the decay is purely spontaneous, leading to the same rate as in an inertial vacuum.

\subsection{Decay Rate in the Intermediate Regime ($a l \sim 1$)}
We now analyze the decay of the cavity's fundamental mode for a cavity of intermediate size, where the acceleration length scale is comparable to the cavity's proper length: $a l \sim 1$. This regime is characterized by the parameter $\beta = \Omega_1/a = (1+s)\pi/\ln(1+al)$ being of order unity. We maintain the condition $M/a \ll 1$ for the external field. These two conditions ensure $M l = (M/a)(a l) \ll 1$. 

The full expression for the squared modulus of the overlap integral, $|f(\Omega_1|^2$, derived from Eqs. \eqref{eq:Mode Overlap Rindler}, contains multiple terms (I, II, III) involving Gamma functions and rapidly oscillating phases. However, for $M/a \ll 1$, a detailed analysis shows that Term I dominates the behavior. The dominant term is:
\begin{align}
|f(\Omega_1)|^2\sim \frac{l^2a^2\pi}{M \cosh(\pi \beta)}\left[1- \sin\left(\theta + 2 \arg\Gamma(1/2 +i\beta) -2\beta \ln\left(\frac{M}{2a}\right)\right) \right]\, .
\end{align}
Terms II and III are proportional to $(M/2a)$ and $1/(\beta^2 + 1/4)$, respectively. For $M/a \ll 1$ and $\beta\sim 1$, these terms are suppressed compared to term I, and are therefore negligible. (A full discussion is provided in Appendix \ref{app:Appendix A}.)

The expression for $|f(\Omega_1)|^2$ contains terms with rapidly oscillating phases proportional to $\ln(M/2a)$. Any infinitesimal experimental uncertainty in the acceleration $a$ or other parameters would cause these terms to average to zero. Therefore, the observable decay rate is governed by the remaining non-oscillatory term. Combing this with the expression \eqref{eq:DecayRate-LeftWallObs_R} yields the averaged decay rate:
\begin{equation}\label{Decay:Probability:al=1}
\Gamma_{\text{left}} \sim \frac{ g^2 l^2 a}{\ln (1 + a l)} \frac{1}{1 + e^{-2\pi\beta}}\, . 
\end{equation}
For $a l \sim 1$, $\beta = \frac{(1+s)\pi}{\ln(1+al)}$. Taking a representative value $s=1/2$, we find $\beta \approx \frac{3\pi}{2\ln 2} \approx 6.8$. An upper bound for $s \in (0, 1)$ is $\tilde{\beta} = \frac{2\pi}{\ln (2)} \approx 9.06$. Thus $e^{-2\pi\beta} \ll 1$ for the entire range of $s$, and the factor $(1 + e^{-2\pi\beta})^{-1} \approx 1$. The averaged decay rate simplifies to:
\begin{align}
\Gamma_{\text{left}} \sim  g^2 l \frac{al}{\ln (1 + a l)} = \Gamma_{\text{in}}\frac{al}{\ln (1 + a l)}\, . \label{eq:finalgamma}
\end{align}

The result includes a logarithmic correction factor \(a l / \ln(1+ a l)\) that arises from the geometric properties of the accelerated cavity. This represents a measurable non-inertial effect distinct from the thermal Unruh signature. In the limit of a very small cavity ($a l \ll 1$), we recover the previous result from Sec.~\ref{sec:ShortCavity al<<1}: $\ln (1+ a l) \approx a l$, which implies $\Gamma_{\text{left}} \sim \Gamma_{\text{in}}$. This expression furthermore converges exactly to the idealized result in the limit $al\rightarrow 0$, where $\lim_{al\rightarrow 0}\frac{al}{\ln (1+ a l)} = 1$, implying $\lim_{al\rightarrow 0} \Gamma_\text{left} = \Gamma_\text{in}$. The expression also contains the fermionic thermal factor \(1/(1 + e^{-2\pi\beta})\), representing Fermi-Dirac stimulation from the Unruh thermal bath. However, for \(a l \sim 1\), this thermal factor is approximately unity, indicating that the thermal Unruh effect is negligible in this parameter regime. The observed enhancement therefore represents primarily geometric/non-thermal acceleration effects, demonstrating that accelerated cavities exhibit measurable deviations from inertial behavior through mechanisms that may be distinct from the thermal Unruh signature. Consequently, while our framework incorporates the effect of the Unruh thermal bath on the decay process, its contribution for $al\sim 1$  is negligible since the enhancement to the decay rate arises primarily from kinematic and geometric effects tied to the cavity's size and acceleration and not from thermal stimulation.  This result provides an important prediction for experimental tests in cavity QED setups with fermionic systems.

\begin{figure}[htbp]
  \centering
  \includegraphics[width=0.8\columnwidth]{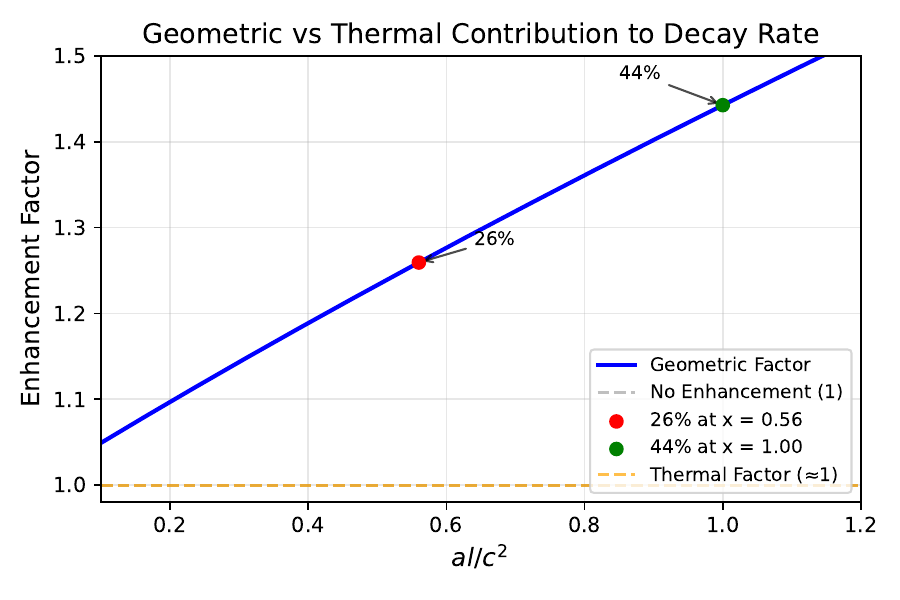}
  \caption{
    The geometric contribution to the decay rate enhancement, $\frac{a l / c^2}{\ln(1 + a l / c^2)}$, 
    as a function of the dimensionless parameter $a l / c^2$. 
    A 26\% enhancement occurs at $a l / c^2 \approx 0.56$ (red marker), corresponding to 
    $a \approx 10^{20} \, \text{m/s}^2$ and $l \approx 500~\mu\text{m}$. 
    The enhancement peaks at $\approx 44\%$ for $a l / c^2 \approx 1$ (green marker). 
    The thermal factor (dashed orange) remains near unity, masking the gray ``No Enhancement (1)" reference line, which is also plotted at x=1.00. The geometric factor alone drives the observed enhancement, dominating over the negligible thermal contribution.
  }
  \label{fig:enhancement}
\end{figure}

The geometric enhancement $\frac{\Gamma_{\text{acc}}}{\Gamma_{\text{in}}} \sim \frac{a l / c^2}{\ln(1 + a l / c^2)}$ peaks near 44\% at $a l / c^2 \approx 1$. For realistic parameters ($a=10^{20}$ m/s$^2$, $l=500~\mu$m $\Rightarrow a l/c^2 \approx 0.56$), it reaches 26\%—a readily measurable signature within experimental constraints. This is illustrated in Fig.~\ref{fig:enhancement}.

\subsection{Decay Rate for a Heavy External Field ($M/a \gg 1$)}
\label{sec:HeavyField}
The previous analyses assumed a light external field ($M/a \ll 1$), which maximizes the response of the Unruh thermal bath. We now consider the opposite regime, where the external field is heavy relative to the acceleration scale ($M/a \gg 1$). In this case, the energy required to excite a quanta of the external field, $Mc^2$, is much larger than the characteristic energy scale of the Unruh thermal bath, $\hbar a/c$. Consequently, we expect the decay process to be dominated by exponential suppression, irrespective of the cavity size. This section analyzes this suppression for small and intermediate cavities ($al \lesssim 1$). 
The analysis for large cavities $(al \gg 1)$ is covered in Case C  of Sec.~\ref{sec:LargeCavity}. 

The derivation of the overlap integral $|f(\Omega_1|^2$ in this regime is involved and is detailed in Appendix~\ref{app:Appendix B}. The key insight is that the asymptotic form of the modified Bessel functions $K_{i\beta \pm 1/2}(M\rho)$ for large $M/\rho$ dominates the integral, leading to a universal exponential factor of $\exp(-2M/a)$. 

\subsubsection*{Case 1: Short Cavity ($al \ll 1$) with $M/a \gg 1$}
For a short cavity, the parameter $\beta =\Omega_1/a \gg 1$ remains large. The detailed derivation in Appendix~\ref{app:Appendix B}  shows that the dominant behavior of the squared modulus of the overlap function \eqref{ModeOverlap:shortCav-Rindler} in this regime takes the form:
\begin{align}
|f(\Omega_1)|^2 = \frac{\pi a^2 l^2 e^{-2M/a}}{M} F_{\text{short}}(a, l, M,s, \theta)\, ,
\end{align}
where the dimensionless algebraic prefactor $F_{\text{short}}$ encapsulates terms of order unity, as derived in Appendix~\ref{app:Appendix B}. Its precise form does not alter the dominant exponential behavior.
Consequently, the decay rate \eqref{eq:DecayRate-LeftWallObs_R} becomes: 
\begin{align}
\Gamma_{\text{left}}\sim \frac{\Gamma_{\text{in}}}{2} \,  e^{\frac{(1+s)\pi^2}{al}}e^{-2M/a} F_{\text{short}}(a, l, M, s,\theta)\, . 
\end{align}

The condition for a precise cancellation of the exponential enhancement and suppression is $(1+s)\pi^2 = 2Mlc/\hbar$ (in SI units). For any fundamental particle (e.g., an electron) and any admissible boundary condition ($s\in(0,1)$), this equation requires a cavity size $l$ on the order of picometers.
For example, with $M = M_e$ (electron mass) and $a =10^{20}\, \text{m/s}^2$, solving $(1+s)\pi^2 = 2Mlc/\hbar$ yields $l \approx 1\text{pm}$, which is  
far smaller than any realistic experimental cavity, which would typically be micrometers or larger. Therefore, the finely-tuned cancellation is physically unattainable. 

This overwhelming dominance is best illustrated with a numerical example. 
Consider a scenario where the cavity undergoes an extreme acceleration of
$a =10^{20}\, \text{m/s}^2$. The energy required to create a quanta of the external electron field  ($M_e c^2 \approx 0.511 \, \text{MeV}$) as the excited state of the confined field decays to its ground state must be compared to the characteristic energy of the acceleration, $\hbar a/c$. The disparity between these energy scales is immense: 
\begin{equation}
\frac{M_e c^2}{\hbar a/c}\approx \frac{0.511 \times 10^6 \text{eV}}{2.2\times 10^{-4}\text{eV}}\approx 2.3\times 10^9\, .
\end{equation}
Now, consider a realistic, small cavity with a length $l = 1 \mu\text{m}$. The enhancement exponent is:
\begin{equation}
\frac{(1+s)\pi^2}{al/c^2} \approx \frac{15}{1.11\times 10^{-3}} =  1.35\times10^4\, ,
\end{equation}
where we have used $al/c^2 \approx 1.11 \times 10^{-3}$ and $(1+s)\pi^2 \approx 15$ (for $s=1/2)$. The net decay rate is proportional to:
\begin{equation}
\exp(1.35\times 10^4)\cdot \exp(-4.6 \times 10^{9}) =  \exp(-4.6 \times 10^{9} + 1.35\times10^4) \approx  \exp(-4.6 \times 10^{9})\, .
\end{equation}
The suppression factor $\exp(-4.6\times 10^{9})$ is so overwhelmingly small that it completely dominates the enhancement factor $\exp(1.35\times 10^4)$, rendering the net rate $\exp(-4.6\times 10^{9})$ utterly immeasurable.
This makes perfect physical sense: creating a massive particle requires so much energy that even with the geometric enhancement from the cavity, the Unruh thermal bath at achievable accelerations cannot provide it,  leading to exponential suppression of the decay process.

Across the entire parameter space, the exponential suppression term $\exp(\frac{-2Mc^2}{\hbar a/c})$  will overwhelmingly dominate the decay rate for any achievable acceleration. This shut-down of the decay channel occurs because the energy required to create a quantum of the external field ($\sim Mc^2$) is vastly larger than the characteristic energy scale of the Unruh thermal bath ($\sim \hbar a/c$),  making the process exponentially improbable. In other words, 
the Unruh thermal bath is too ``cold"  ($k_B T_U = \hbar a/(2\pi c)$) to excite the heavy external field, shutting down the decay channel.

\subsubsection*{Case 2: Intermediate Cavity ($al\sim 1$) with $M/a \gg 1$}
In this regime, $\beta =\Omega_1/a = (1+s)\pi/\ln(1+al)$ is of order unity (since $s\in(0,1)$ and ($al\sim 1$). Despite this change, the condition $M/a\gg 1$
ensures the exponential suppression remains the dominant effect. The analysis yields a result of the same form:
\begin{align}
|f(\Omega_1)|^2 = \frac{\pi a^2 l^2 e^{-2M/a}}{M} G_{\text{short}}(a, l, M,s, \theta)\, ,
\end{align}
where the dimensionless algebraic prefactor $G_{\text{short}}$ is given in Appendix~\ref{app:Appendix B}. Consequently, the decay rate \eqref{eq:DecayRate-LeftWallObs_R} becomes:
\begin{align}
\Gamma_{\text{left}}\sim \frac{\Gamma_{\text{in}}}{2}\frac{al}{\ln(1+al)} \,  e^{\frac{(1+s)\pi^2}{\ln(1+al)}}e^{-2M/a} G_{\text{short}}(a, l, M,s, \theta)\, . 
\end{align}
While the algebraic prefactors $F_{\text{short}}$ and $G_{\text{short}}$ differ from the short-cavity case, the governing exponential factor is identical. The enhancement exponent $(1+s)\pi^2/\ln(1+al)\approx(1+s)\pi^2/\ln 2$ (since $al\sim 1$) is vastly smaller than the suppression exponent $2M/a = 4.6\times 10^{9}$ (where $2M/a$ corresponds to $2Mc^2/(\hbar a/c)$ in SI units). Thus again, the exponential suppression overwhelmingly dominates the decay rate.

This analysis confirms that for a heavy external field
($M/a \gg 1$), the decay rate is exponentially suppressed across all cavity sizes, $\Gamma_{\text{left}}/\Gamma_{\text{in}}\sim e^{-2M/a}$. The result derived for large cavities in Sec.~\ref{sec:LargeCavity} (Case C) is not an isolated case but is, in fact, the universal behavior for this parameter regime. The specific cavity size $al$ only influences the algebraic prefactor, which is always many orders of magnitude smaller than the exponential factor $\exp(-2M/a)$, 
 but not the dominant exponential character of the suppression.
 This suppression may help explain why experiments utilizing electrons or other massive fundamental particles—which \(Mc^2/(\hbar a/c) \gg 1\)—have not observed Unruh effects.

\subsection{Decay Rate in the Large Cavity Regime ($al \gg 1$)}
\label{sec:LargeCavity}

We now analyze the decay of the cavity's fundamental mode in the regime where its proper length is large compared to the acceleration length scale, $al \gg 1$. This implies the parameter $\beta = \Omega_1/a = (1+s)\pi/\ln(1+al)$ is small ($\beta \ll 1$),  as the logarithmic denominator is large. In this regime, the proper acceleration varies little across the cavity, and we expect the influence of the Unruh effect to diminish. The decay dynamics are governed by the secondary parameter $M/a$, the ratio of the external field's mass to the acceleration, leading to three distinct physical subcases. We use the approximations $\ln(1+al) \approx \ln(al)$ and $e^{\pi \beta} \approx 1$. 

The overlap integral \eqref{eq:Mode Overlap Rindler} for $f(\Omega_1)$ must be evaluated by integrating over $\rho$. To leading order in the small parameter $\beta$ ($(O(\beta^0))$, it simplifies to:
\begin{equation}
f^{(0)}(\Omega_1) \approx (e^{i\theta} - i) \int_{1/a}^{1/a + l} \frac{d\rho}{\sqrt{\rho}} K_{\frac{1}{2}}(M\rho)\, . \label{eq:f0_integral}
\end{equation}
The behavior of this integral defines the following subcases. (For a full discussion of the asymptotic analysis see Appendix \ref{app:Appendix C}.)

\subsubsection{Case A: Light External Field ($M/a \ll 1$); implies $Ml \ll 1$)}
Here, the argument of the Bessel function $M\rho$ is small. Using $K_{1/2}(z)\sim \sqrt{\pi/(2z)}=K_{-1/2}(z)$ for $z\ll 1$,  we obtain for squared modulus of the overlap function:
\begin{align}
|f^{(0)}(\Omega_1)|^2 \sim \frac{\pi}{M}(1 - \sin\theta) \ln^2(a l) \, . \label{eq:f2_A}
\end{align}
Collecting \eqref{eq:DecayRate-LeftWallObs_R} yields the decay rate:
\begin{equation}
\Gamma_{\text{left}} \sim \frac{g^2}{2a} (1 - \sin\theta) \ln(a l)\, . \label{eq:Gamma_A}
\end{equation}
This has a clear physical interpretation: The decay rate  scales algebraically with the acceleration ($\Gamma_{\text{left}} \propto 1/a$) with logarithmic corrections. This signifies a return to inertial-like behavior, modulated by the boundary condition through $(1-\sin\theta)$. 

\subsubsection{Case B: Heavy  Field -- Light Relative to Acceleration ($M/a \ll 1$ but $Ml \gg 1$)}

This case explores the regime where $M/a \ll 1$ (so the field is light relative to acceleration) but $Ml \gg 1$
(so the field is heavy relative to the cavity length). This requires $al \gg a/M$, which is possible since $a/M \gg 1$. Using
$K_{1/2}(z)\sim \sqrt{\pi/(2z)}$  and the exponential integral $E_1(M/a)= \int_{1/a}^\infty \frac{e^{-M\rho}}{\rho}d\rho \approx -\gamma -\ln (M/a)$ for $M/a \ll 1$, we find
for the squared modulus of the overlap integral:
\begin{equation}
|f^{(0)}(\Omega_1)|^2 \approx \frac{\pi}{M}(1 - \sin\theta) \left( \ln(a/M) - \gamma \right)^2 \, . \label{eq:f2_B}
\end{equation}
where $\gamma$ is the Euler-Mascheroni constant. Collecting \eqref{eq:DecayRate-LeftWallObs_R}, the resulting decay rate is:
\begin{equation}
\Gamma_{\text{left}} \sim \frac{g^2}{2a \ln(a l)} (1 - \sin\theta) \left( \ln(a/M) - \gamma \right)^2. \label{eq:Gamma_B}
\end{equation}
The rate is further suppressed by the factor  $1/\ln(al)$ compared to Case A, reflecting the increased difficulty of exciting a heavier field, even when $M/a \ll 1$.

\subsubsection{Case C: Heavy External Field -- Strong Relative to Acceleration ($M/a \gg 1$; implies $Ml \gg 1$)}

We now consider the case where the external field's mass is large compared to the acceleration scale, $M/a \gg 1$. The argument $z =M\rho$ is large, requiring the large-argument asymptotic expansion $K_\nu \sim \sqrt{\pi/(2z)} e^{-z}$. The integral in \eqref{eq:f0_integral} is again dominated by its lower limit. Using the Laplace method for integrals of the form $\int e^{-M\rho}g(\rho) d\rho$ with $g(\rho) =1/\rho$ varying slowly compared to the exponential, we obtain 
for the squared modulus of the overlap function:
\begin{equation}
|f^{(0)}(\Omega_1)|^2 \approx \frac{\pi a^2}{2 M^3} |e^{i\theta} - i|^2 e^{-2M/a} = \frac{\pi a^2}{M^3} (1 - \sin\theta) e^{-2M/a}. \label{eq:f2_C}
\end{equation}
Collecting \eqref{eq:DecayRate-LeftWallObs_R} yields the decay rate: 
\begin{equation}
\Gamma_{\text{left}} \sim \frac{g^2 a}{2M^2 \ln(a l)} (1 - \sin\theta) e^{-2M/a}. \label{eq:Gamma_C}
\end{equation}
The decay rate is exponentially suppressed by the factor $e^{-2M/a}$. This confirms that the Unruh thermal bath at temperature $T=a/(2\pi)$ is too "cold" to excite particles of mass $M \gg a$. For the specific case of the MIT bag model $(\theta = \pi/2)$, the prefactor $(1-\sin\theta) =0$, causing the zeroth-order term to vanish exactly. The decay would then be governed by higher-order terms in $\beta$, which are also exponentially suppressed.

\subsubsection{Summary}
The behavior of the decay rate in the large cavity regime ($ al\gg 1$) transitions from inertial-like to exponentially suppressed, depending on the mass of the external field relative to the acceleration. Table \ref{tab:summary_large_cavity} summarizes the key results.
\begin{table}[h!]
\centering
\caption{Summary of decay rate $\Gamma_{\text{left}}$ behavior in the large cavity regime ($a l \gg 1$).}
\label{tab:summary_large_cavity}
\adjustbox{max width=\linewidth}{ 
\begin{tabular}{c l l l}
\toprule
\textbf{Case} & \textbf{Condition} & \textbf{Decay Rate $\mathbf{\Gamma}_{\text{left}}$} & \textbf{Dominant Behavior} \\
\midrule
\textbf{A} & $M/a \ll 1$ & $\dfrac{\Gamma_{\text{in}}}{2}\dfrac{(1 - \sin\theta)\ln(a l)}{a l} $ & Algebraic ($\sim 1/a$) \\
    & (implies $M l \ll 1$) & & \\
\midrule
\textbf{B} & $M/a \ll 1$, $M l \gg 1$ & $\dfrac{\Gamma_{\text{in}}}{2}\dfrac{(1 - \sin\theta) \left( \ln(a/M) - \gamma \right)^2}{a l \ln(a l)}$ & Algebraic, logarithmically suppressed \\
\midrule
\textbf{C} & $M/a \gg 1$ & $\dfrac{\Gamma_{\text{in}}}{2}\dfrac{(1 - \sin\theta) e^{-2M/a}}{{(M/a)(M l) \ln(a l)} }$ & Exponentially suppressed ($\sim e^{-2M/a}$) \\
    & (implies $M l \gg 1$) & & \\
\bottomrule
\end{tabular}
} 
\end{table}

\section{Conclusion}
\label{sec:conclusion}

Our analysis demonstrates that the decay rate of an excited state within a uniformly accelerated cavity depends on two key dimensionless parameters: the cavity size relative to the acceleration scale, $al$, and the mass of the external field relative to the acceleration, $M/a$. For light external fields $(M/a \ll 1)$, the decay rate transitions from inertial-like behavior ($\Gamma_{\text{left}} \sim \Gamma_{\text{in}}$) for $al \ll 1$, to a geometric enhancement ($\Gamma_{\text{left}} \sim \Gamma_{\text{in}}\frac{al}{\ln(1+al)}$) for $al \sim 1$, and to an algebraically suppressed rate ($\Gamma_{\text{left}} \sim \Gamma_{\text{in}}/al$) with logarithmic corrections for $al \gg 1$. For heavy external fields ($M/a \gg 1$), the decay rate is exponentially suppressed ($\Gamma_{\text{left}}/\Gamma_{\text{in}} \sim e^{-2M/a}$) regardless of the cavity size. These results highlight the rich interplay between acceleration, cavity size, and field mass in quantum field theory.

Within the context of our model, the exponential suppression for heavy fermionic fields could help explain why Unruh-induced modifications to particle decay rates have not been observed for heavy fermions. For fundamental fermionic particles like electrons, the condition $Mc^2 \gg \hbar a/c$ is satisfied for all achievable accelerations. For instance, an electron ($M_ec^2 \approx 0.511$ MeV) at an acceleration of $a =10^{20}$ m/s$^2$ ($\hbar a/c \approx 2.2\times 10^{-4}$ eV) yields a suppression factor of $\Gamma_{\text{acc}}/\Gamma_{\text{in}} \sim \exp(-4.6 \times 10^9)$. This enormous disparity in energy scales — between the fermion’s rest mass and the Unruh scale $\hbar a/c$ — results in exponential suppression that overwhelms any potential Unruh-induced signal in decay processes involving heavy fermions.

This analysis demonstrates that no foreseeable improvement in acceleration technology can overcome this suppression within the framework of our model. The suppression scales as $\Gamma_{\text{acc}}/\Gamma_{\text{in}} \sim \exp(-C/a)$ where $C = 2Mc^3/\hbar$. For electrons, this gives $C \approx 4.6\times 10^{29}$ m/s$^2$. At an acceleration of $a = 10^{20} \, \text{m/s}^2$, this yields an exponent $C/a \approx 4.6 \times 10^{9}$. Increasing acceleration by a factor $K$ yields $\Gamma_{\text{acc}} \sim \exp(-C/(Ka))$, requiring $K \gtrsim 4.6\times 10^9$ for measurability—physically unrealizable.

Quantum simulators circumvent this issue by engineering effective light fields where the effective mass energy $M_{\text{eff}}c^2$ can be made very small (e.g., through low-frequency modes or synthetic materials). This allows the condition $M_{\text{eff}}c^2 \ll \hbar a/c$ to be satisfied, enabling access to a regime where the decay rate is governed by a measurable, algebraic enhancement. These platforms provide the most promising ways for detecting Unruh-like effects in decay processes accompanied by excitation of massive field quanta.

Consequently quantum simulation platforms, particularly superconducting circuits, provide the most viable path for  testing  our model's prediction of the geometric enhancement $\Gamma_{\text{left}}/\Gamma_{\text{in}}\sim \frac{al/c^2}{\ln (1+al/c^2)}$ for an intermediate cavity size ($al\sim c^2$). Crucially, for the external environment to simulate the massive Dirac field, the effective mass scale must satisfy $M_{\text{eff}}c^2 \ll \hbar a/c$, which for typical acceleration parameters ($a = 10^{20} \, \text{m/s}^2$) requires an effective mass energy below $10^{-4}\text{eV}$. This can be achieved by engineering a low-frequency environment or using synthetic materials. In this architecture, micrometer-scale coplanar waveguide resonators ($l\sim 100-500 \mu\text{m}$) are ideally suited to pursue the $al\sim c^2$ regime. The required effective acceleration  can be simulated via time-dependent modulation of circuit parameters at microwave frequencies ($\sim 10-100 \text{GHz}$).

 The beam-splitter interaction Hamiltonian $H_\text{int}=\int d\omega g(\omega)(\hat{a}\hat{b}_\omega^\dagger + \hat{a}^\dagger\hat{b}_\omega)$ naturally implements the local coupling between the cavity mode (simulating the massless Dirac field, represented by operator $\hat{a}$) and the external continuum (simulating the massive Dirac field, represented by operator $\hat{b}_\omega$). We note that the massless Dirac field in our model is characterized by a linear dispersion relation, while the massive Dirac field has a dispersion relation with a mass gap. In superconducting circuits, we can engineer systems that have these dispersion relations. The confined field (massless) can be a microwave resonator with a linear dispersion. To capture the massive nature of the external field, the continuum must have a bandgap, such that the density of states is zero for frequencies below a gap $\Delta$ and continuous above. This can be engineered in superconducting circuits by using a photonic crystal waveguide or a transmission line coupled to a qubit, which creates a gap of size $\Delta$. The condition for decay is that the cavity frequency $\Omega_1$ must be above the gap $(\Omega_1 >\Delta)$, and the effective mass is given by $M_\text{eff}=\Delta/c^2$. The direct energy-conserving decay process $\hat{a}\hat{b}_\omega^\dagger$—where a photon in the cavity mode is converted to an excitation in the external continuum above the gap—faithfully reproduces the core physics of our model, while the massless/massive field distinction is encoded in the engineered dispersion relations and density of states. The experimental protocol involves first measuring the inertial decay rate  $\Gamma_\text{in}$
 as a baseline under the condition$Mcl/\hbar \ll 1$ (or $Ml\ll 1$ in natural units), and then comparing it to the enhanced rate $\Gamma_\text{acc}$ under effective acceleration engineered via time-dependent parameter modulation.

We emphasize that while our theoretical model involves fermionic fields, the quantum simulation uses bosonic fields (microwave photons) to simulate the essential physics, particularly the coupling between a discrete state and a continuum and the resulting decay dynamics. The beam-splitter interaction Hamiltonian captures the bilinear coupling of the model, and the statistics of the fields are not expected to alter the single-particle decay process we study.

Notwithstanding the possibility of testing this model, our work has several inherent limitations. The analysis is conducted entirely within a (1+1)-dimensional spacetime  framework, which neglects key physical features present in realistic (3+1)-dimensional settings, such as the absence of transverse spatial degrees of freedom, limited mode structure, and the lack of angular momentum or helicity effects. Furthermore, the model considers a massless confined field coupled to a massive external field — a specific configuration chosen for analytical clarity that may not capture all dynamics of systems where both fields are massive. As a result, care must be taken in extrapolating our conclusions — particularly the geometric enhancement $\Gamma_{\text{left}}/ \Gamma_{\text{in}} \sim \frac{al}{\ln(1+al)}$) for $al \sim 1$— to higher-dimensional experimental implementations.

Despite these theoretical simplifications, we emphasize that the origin of this enhancement lies in the asymptotic properties of the mode overlap integrals, which involve Modified Bessel functions whose large-argument behavior is dimension-independent. This suggests that the qualitative phenomenon — a non-thermal, kinematic enhancement of the decay rate due to acceleration and cavity size — is likely robust and may generalize to 3+1D, albeit with a different functional form for the geometric factor $F_g$. Indeed, we hypothesize that in an identical regime, a similar factorization $\Gamma_{\text{acc}}/ \Gamma_{\text{in}} \sim F_g F_T$ holds for other field configurations and other choices of boundary conditions, though $F_g$ may differ from the 1+1D expression derived here.
Similarly, the exponential suppression
 $\Gamma_{\text{acc}}/\Gamma_{\text{in}}  \sim \exp(-M/a)$ for heavy external fields is expected to persist in higher dimensions, as it stems from the energy-scale mismatch $M  \gg  a $, a condition independent of spacetime dimensionality. While the specific algebraic prefactor could vary with different field types, couplings, or cavity conditions, the dominance of the exponential suppression when $M  \gg a $ is likely a universal physical mechanism that is expected to hold for any realistic detection scheme relying on the excitation of heavy massive field quanta.

Future work should aim to extend this analysis to (1+3)-dimensions and explore more general field configurations to assess the dominance of geometrical enhancement over thermal stimulation from the Unruh thermal bath as well as the reported exponential suppression for heavy fermionic particle.

\section*{Acknowledgments}

We thank Wan Mohamad Husni Wan Mokhtar for helpful discussions. We also thank Jorma Louko for helpful discussions. 
\appendix

\section{Full Asymptotic Expression and its Simplification}
\label{app:Appendix A}
This appendix contains a complete derivation of the asymptotic expression for the squared modulus of the overlap integral, given in  \eqref{eq:OverlapMagnitude}. This result justifies the simplification used in the main text to obtain the decay rate.

Starting from the approximation for the overlap integral \eqref{ModeOverlap:shortCav-Rindler}, valid under the condition $Ml \ll 1$, 
\begin{align}\label{ModeOverlap:shortCav-Rindler:al=1}
f(\Omega_1) \approx l\sqrt{a}\left[e^{i\theta}K_{i\beta +\frac{1}{2}}(M/a) -i K_{i\beta -\frac{1}{2}}(M/a)\right]\, .
\end{align} 
For $z\rightarrow 0$ and fixed $\nu$, the Modified Bessel function of the second kind has the asymptotic expansion  \cite{DLMF}: $K_\nu(z)\sim \frac{1}{2}\Gamma(-\nu) \left(\frac{z}{2}\right)^\nu + \frac{1}{2}\Gamma(\nu) \left(\frac{z}{2}\right)^{-\nu}$.
Applying this to the term in \eqref{ModeOverlap:shortCav-Rindler:al=1} with $z= M/a\ll 1$ gives:
\begin{align}
K_{i\beta \pm \frac{1}{2}} \sim \frac{1}{2}\Gamma(-i\beta \mp \frac{1}{2})\left(\frac{z}{2} \right)^{i\beta \pm \frac{1}{2}} +  \frac{1}{2}\Gamma(i\beta \pm \frac{1}{2})\left(\frac{z}{2} \right)^{-i\beta \mp \frac{1}{2}} \, .
\end{align}
After substituting the asymptotic expansions and simplifying, we obtain:
\begin{align}
|f(\Omega_1)|^2 = \frac{l a}{4} \left( \text{I} + \text{II} + \text{III} \right)\, ,
\end{align}
where the terms $\text{I}$, $\text{II}$ and $\text{III}$ are defined as follows:
\begin{align}
\text{I} &= \left( \frac{z}{2} \right)^{-1} \left[ \frac{2\pi}{\cosh(\pi \beta)} + 2 \Re \left( i e^{i\theta} \,\Gamma(i\beta + 1/2)^2 \left( \frac{z}{2} \right)^{-2i\beta} \right) \right]\, , \\
\text{II} &= \frac{z}{2} \left[ \frac{2\pi}{(\beta^2 + 1/4) \cosh(\pi \beta)} + 2 \Re \left( i e^{i\theta} \,\Gamma(-i\beta - 1/2)^2 \left( \frac{z}{2} \right)^{2i\beta} \right) \right] \, ,\\
\text{III} &= -\frac{2}{\beta^2 + 1/4} \Re \left( (1/2 - i\beta) \,\Gamma(1/2 - i\beta)^2 \left( \frac{z}{2} \right)^{2i\beta} \right) \notag \\ &+ \frac{2\pi}{\cosh(\pi \beta)} \frac{ \sin\theta - 2\beta \cos\theta}{\beta^2 + 1/4}\, .
\end{align}
Here $\beta = \frac{\Omega_1}{a}=\frac{\pi(1+s)}{\ln(1+al)}$, and $\Gamma(z)^2$ stands for the square of the Gamma function (not the squared modulus). This expression, while exact within the asymptotic approximation, is complex due to the oscillatory phase $(z/2)^{\pm 2i\beta}$. However, the physical regime of interest $z=(M/a)\ll 1, al\sim 1$ allows for a significant simplification.

 Notably, the full expression for $|f(\Omega_1)|^2$ contains terms with coefficients of order $(z/2)^{-1}, (z/2)^0$, and $(z/2)^1$. Term I is proportional to $(z/2)^{-1}$. This is the dominant contribution. Term II is proportional to $(z/2)^1$. Since $z \ll 1$, this term is suppressed by a factor of $z^2$ compared to Term I and is negligible. Term III is proportional to $(z/2)^0$. Although it is not parametrically small like Term II, its coefficient is subdominant. For the values of $\beta\sim 1$ relevant for $al \sim 1$, the prefactors in Term III ($1/(\beta^2 +1/4)\sim 1$) are order unity. Note, however, the entire Term III is significantly smaller than the the constant part of Term I.  For example, for the specific case of the MIT bag model $(\theta =\pi/2)$, the non-oscillatory part of Term III simplifies to $\frac{2\pi}{\cosh(\pi \beta)}\frac{1}{\beta^2 +1/4}$, which remains subdominant. The oscillator terms of both Term I and Term III, such as $\Re \left( i e^{i\theta} \,\Gamma(i\beta + 1/2)^2 \left( \frac{z}{2} \right)^{-2i\beta} \right)$  and $\Re \left((1/2-i\beta  ) \Gamma(i\beta - 1/2)^2\left( \frac{z}{2} \right)^{2i\beta} \right)$ have amplitudes of order unity but oscillate extremely rapidly as a function of $\beta$ (and thus of the acceleration $a$) due to the factor $(z/2) = \exp[\pm 2i\beta\ln(z/2)]$. Any infinitesimal experimental uncertainty in the acceleration $a$ would cause these terms to average to zero upon measurement. Therefore, only the non-oscillatory, constant terms contribute to the measurable average.
 
Consequently, to an excellent approximation, we retain only the constant part of the dominant Term I:
\begin{align}
|f(\Omega_1|^2 \sim \frac{l^2 a \pi}{z \cosh(\pi \beta)} = \frac{ l^2a^2 \pi}{M\cosh(\pi \beta)}\, .
\end{align}
This is the simplified form used in Eq.\eqref{Decay:Probability:al=1} of the main text to compute the decay probability.

\section{Asymptotic Analysis of the Overlap integral for $M/a \gg 1$}
\label{app:Appendix B}

In this appendix we analyze the asymptotic behavior of the overlap integral \eqref{eq:Mode Overlap Rindler} 
in the regimes where $al \ll 1$ and $al\sim 1$ for $M/a \gg 1$. 

The overlap integral is defined as:
\begin{equation}
f(\Omega_1) = \int_{\rho_-}^{\rho_- + l} \frac{d\rho}{\sqrt{\rho}} \left[ e^{i\theta} \left( \frac{\rho}{\rho_-} \right)^{-i\frac{\Omega_1}{a}} K_{i\frac{\Omega_1}{a} + \frac{1}{2}}(M\rho) - i \left( \frac{\rho}{\rho_-} \right)^{i\frac{\Omega_1}{a}} K_{i\frac{\Omega_1}{a} - \frac{1}{2}}(M\rho) \right],
\label{eq:main_integral}
\end{equation}
where $\rho_- = 1/a$, $\beta = \Omega_1 / a$, and $K_\nu(z)$ is the modified Bessel function of the second kind. For $M/a \gg 1$, the argument $M\rho$ satisfies $M\rho \geq M/a \gg 1$ throughout the integration range, allowing us to use the large-argument expansion:
$
K_\nu(z) \approx \sqrt{\frac{\pi}{2z}} e^{-z}.
$

Changing variables to $u = \rho - \rho_-$ with $d\rho = du$ and limits from $0$ to $l$, we obtain the general asymptotic form:
\begin{equation}
f(\Omega_1) \approx \sqrt{\frac{\pi}{2M}} \int_{0}^{l} \frac{du}{\rho_- + u} e^{-M(\rho_- + u)} \left[ e^{i\theta} \left( \frac{\rho_- + u}{\rho_-} \right)^{-i\beta} - i \left( \frac{\rho_- + u}{\rho_-} \right)^{i\beta} \right]\, .
\label{eq:general_form_step1}
\end{equation}
We now analyze this integral in two distinct regimes.

\section*{Case 1: Short Cavity ($a l \ll 1$)}
When $a l \ll 1$ (i.e., $l \ll 1/a$), we have $u \ll \rho_- = 1/a$ throughout the integration range. This allows the following approximations:
\begin{align}
\frac{1}{\rho_- + u} &\approx \frac{1}{\rho_-} = a\, , \\
\left( \frac{\rho_- + u}{\rho_-} \right)^{\pm i\beta} &= (1 + a u)^{\pm i\beta} \approx e^{\pm i\beta a u} = e^{\pm i\Omega_1 u}, \\
e^{-M(\rho_- + u)} &= e^{-M/a} e^{-M u}\, .
\end{align}
Substituting these approximations into \eqref{eq:general_form_step1} yields:
\begin{equation}
f(\Omega_1) \approx a \sqrt{\frac{\pi}{2M}} e^{-M/a} \int_{0}^{l} e^{-M u} \left[ e^{i\theta} e^{-i\Omega_1 u} - i e^{i\Omega_1 u} \right] du\, .
\label{eq:case1_general}
\end{equation}
This integral can be evaluated exactly, yielding:
\begin{equation}
f(\Omega_1) \approx a \sqrt{\frac{\pi}{2M}} e^{-M/a} \left[ \frac{e^{i\theta}(1 - e^{-(M + i\Omega_1)l})}{M + i\Omega_1} - \frac{i(1 - e^{-(M - i\Omega_1)l})}{M - i\Omega_1} \right]\, .
\label{eq:case1_exact}
\end{equation}
The corresponding squared modulus is:
\begin{align}
|f(\Omega_1)|^2 &\approx \frac{\pi a^2 e^{-2M/a}}{2M} \left| \frac{e^{i\theta}(1 - e^{-(M + i\Omega_1)l})}{M + i\Omega_1} - \frac{i(1 - e^{-(M - i\Omega_1)l})}{M - i\Omega_1} \right|^2\, , \\
&\equiv  \frac{\pi a^2 l^2 e^{-2M/a}}{M} F_{\text{short}}(a, l, M,s, \theta) \, ,
\label{eq:case1_squared}
\end{align}
where the dimensionless prefactor $F_{\text{short}}$ is defined as 
\begin{align}
F_{\text{short}}(a, l, M,s, \theta)\equiv \frac{1}{2l^2} \left| \frac{e^{i\theta}(1 - e^{-(M + i\Omega_1)l})}{M + i\Omega_1} - \frac{i(1 - e^{-(M - i\Omega_1)l})}{M - i\Omega_1} \right|^2\, .
\end{align}
We now consider two subcases based on the value of $M l$.

\subsection*{Subcase 1a: $M l \ll 1$}

When $M l \ll 1$, we expand the exponentials in the integrand of \eqref{eq:case1_general}:
\begin{align}
e^{-M u} &\approx 1 - M u + \mathcal{O}((M u)^2), \\
e^{\pm i\Omega_1 u} &\approx 1 \pm i\Omega_1 u + \mathcal{O}((\Omega_1 u)^2)\, .
\end{align}

Substituting into \eqref{eq:case1_general} and evaluating the integral yields:
\begin{align}
f(\Omega_1) \approx  a \sqrt{\frac{\pi}{2M}} e^{-M/a} \left[ (e^{i\theta} - i) l + \frac{l^2}{2}\left(- i\Omega_1 (e^{i\theta} + i)  - M(e^{i\theta} - i)\right)  \right]\, . \label{eq:case1a_result}
\end{align}

The squared modulus is:
\begin{equation}
|f(\Omega_1)|^2 \approx \frac{\pi a^2 e^{-2M/a}}{2M} \left| (e^{i\theta} - i) l + \frac{l^2}{2}\left(- i\Omega_1 (e^{i\theta} + i)  - M(e^{i\theta} - i)\right)  \right|^2 \, . \label{eq:case1a_squared}
\end{equation}

\subsection*{Subcase 1b: $M l \gg 1$}

When $M l \gg 1$, the exponential $e^{-M u}$ decays rapidly, allowing us to extend the upper limit to infinity:
\begin{equation}
\int_{0}^{l} e^{-M u} \left[ e^{i\theta} e^{-i\Omega_1 u} - i e^{i\Omega_1 u} \right] du \approx \int_{0}^{\infty} e^{-M u} \left[ e^{i\theta} e^{-i\Omega_1 u} - i e^{i\Omega_1 u} \right] du\, .
\end{equation}
Combining with \eqref{eq:case1_general}, we obtain for the overlap integral:
\begin{equation}
f(\Omega_1) \approx a \sqrt{\frac{\pi}{2M}} e^{-M/a} \left[ \frac{e^{i\theta}}{M + i\Omega_1} - \frac{i}{M - i\Omega_1} \right]\, . \label{eq:case1b_result}
\end{equation}
The squared modulus is:
\begin{align}
|f(\Omega_1)|^2 &\approx \frac{\pi a^2 e^{-2M/a}}{2M} \frac{|M(e^{i\theta} - i) + \Omega_1(1 - i e^{i\theta})|^2}{(M^2 + \Omega_1^2)^2}\, , \\
&\equiv  \frac{\pi a^2 l^2 e^{-2M/a}}{M} G_{\text{short}}(a, l, M,s, \theta) \, ,  \label{eq:case1b_squared}
\end{align}
where the dimensionless prefactor $G_{\text{short}}$ is defined as 
\begin{align}
G_{\text{short}} =\frac{1}{2l^2} \frac{|M(e^{i\theta} - i) + \Omega_1(1 - i e^{i\theta})|^2}{(M^2 + \Omega_1^2)^2}\, .
\end{align}
We have verified that the right-hand side has the correct dimensions of $[M]^{-1}$.

\subsection*{Case 2: Intermediate Cavity ($a l \sim 1$)}

When $a l \sim 1$ (i.e., $l \sim 1/a$), we have $M l = (M/a)(a l) \gg 1$ since $M/a \gg 1$. The exponential factor $e^{-M u}$ decays rapidly, allowing us to extend the upper limit to infinity:
\begin{equation}
\int_{0}^{l} e^{-M u} \left[ e^{i\theta} e^{-i\Omega_1 u} - i e^{i\Omega_1 u} \right] du \approx \int_{0}^{\infty} e^{-M u} \left[ e^{i\theta} e^{-i\Omega_1 u} - i e^{i\Omega_1 u} \right] du\, .
\end{equation}
Evaluating these integrals yields:
\begin{equation}
f(\Omega_1) \approx a \sqrt{\frac{\pi}{2M}} e^{-M/a} \left[ \frac{e^{i\theta}}{M + i\Omega_1} - \frac{i}{M - i\Omega_1} \right]\, .
\label{eq:case2_result}
\end{equation}
The squared modulus is given by \eqref{eq:case1b_squared}.

\subsection*{Higher-Order Corrections and Subdominant Behavior}

The asymptotic expressions derived above represent the leading-order behavior of the overlap integral. Higher-order corrections arise from several sources.

\subsection*{Next-to-Leading Order in Bessel Function Expansion}

The large-argument expansion of the modified Bessel function has subleading terms \cite{DLMF}:
\begin{equation}
K_\nu(z) = \sqrt{\frac{\pi}{2z}} e^{-z} \left[ 1 + \frac{4\nu^2 - 1}{8z} + \mathcal{O}(z^{-2}) \right]\, .
\end{equation}
For our case with $\nu = i\beta \pm \frac{1}{2}$, the first correction term is of order:
\begin{equation}
\frac{4(i\beta \pm \frac{1}{2})^2 - 1}{8M\rho} = \frac{-4\beta^2 \pm 4i\beta}{8M\rho} = \mathcal{O}\left(\frac{\beta}{M\rho}\right)\, .
\end{equation}
Since $M\rho \geq M/a \gg 1$ and $\beta = \Omega_1/a$ is typically $\mathcal{O}(1)$ or smaller, these corrections are suppressed by at least $\mathcal{O}((M/a)^{-1})$ relative to the leading order.

\subsection*{Approximation of $(1 + a u)^{\pm i\beta}$}

The approximation $(1 + a u)^{\pm i\beta} \approx e^{\pm i\Omega_1 u}$ has corrections of order:
\begin{equation}
(1 + a u)^{\pm i\beta} = \exp\left[\pm i\beta \ln(1 + a u)\right] = \exp\left[\pm i\Omega_1 u \mp i\beta \frac{a^2 u^2}{2} + \mathcal{O}((a u)^3)\right] \, .
\end{equation}
For $a l \ll 1$, we have $a u \leq a l \ll 1$, so these corrections are suppressed by $\mathcal{O}((a l)^2)$. For $a l \sim 1$, the exact integral evaluation in \eqref{eq:case2_result} inherently accounts for these corrections.

\subsection*{Extension of Integration Limits}

The extension of the upper integration limit from $l$ to $\infty$ in Case 2 introduces an error of order:
\begin{equation}
\int_{l}^{\infty} e^{-M u} \left[ e^{i\theta} e^{-i\Omega_1 u} - i e^{i\Omega_1 u} \right] du \sim e^{-M l} = e^{-(M/a)(a l)}\, .
\end{equation}
Since $M/a \gg 1$ and $a l \sim 1$, this error is exponentially small and thus subdominant.

\subsection*{Summary}

The original integral \eqref{eq:main_integral} has dimension: $[f] = [l][a^{1/2}] = [M^{-1}][M^{1/2}] = [M^{-1/2}]$, since $l$ has dimension $[M^{-1}]$ and $a$ has dimension $[M]$. 
The results maintain dimensional consistency throughout and demonstrate that for $M/a \gg 1$, in both the short-cavity regime ($a l \ll 1$) and intermediate cavity regime ($a l \sim 1$), the overlap integral exhibits exponential suppression: $f(\Omega_1) \sim e^{-M/a} \times (\text{prefactor})$. Higher-order corrections are shown to be subdominant in all cases, suppressed by factors of $(M/a)^{-1}$, $(a l)^2$, or exponentially small terms. 


\section{Asymptotic Analysis of the Overlap Integral for $al \gg 1$}
\label{app:Appendix C}
This appendix details the asymptotic behavior of the overlap integral \eqref{eq:Mode Overlap Rindler} when $al \gg 1$ ($\beta:=\Omega_1/a \ll 1$). The analysis is presented for two subcases: Case A for which $M l \ll 1$ and Case C for which $M/a \gg 1$ (large mass relative to acceleration). 

We begin by rewriting the overlap integral \eqref{eq:Mode Overlap Rindler} for a left-wall observer located at $\rho_-= 1/a$:
\begin{equation}
f = \int_{1/a}^{1/a + l} \frac{d\rho}{\sqrt{\rho}} \left[ e^{i\theta} (a\rho)^{-i\beta} K_{i\beta + \frac{1}{2}}(M\rho) - i (a\rho)^{i\beta} K_{i\beta - \frac{1}{2}}(M\rho) \right]\, .
\label{eq:main_integral}
\end{equation}

We expand in powers of $\beta$ using
\begin{align}
(a\rho)^{\pm i\beta} &= e^{\pm i\beta \ln(a\rho)} = 1 \pm i\beta \ln(a\rho) - \frac{\beta^2}{2} \ln^2(a\rho) + \cdots\, , \label{eq:exp_expansion} \\
K_{i\beta \pm \frac{1}{2}}(z) &= K_{\frac{1}{2}}(z) \pm i\beta \left. \frac{\partial K_\nu}{\partial \nu} \right|_{\nu=\frac{1}{2}} - \frac{\beta^2}{2} \left. \frac{\partial^2 K_\nu}{\partial \nu^2} \right|_{\nu=\frac{1}{2}} + \cdots \, . \label{eq:bessel_expansion}
\end{align}
Using the symmetry $K_{-\nu}(z) = K_\nu(z)$ (with $z = M\rho$), we obtain:
\begin{equation}
f = f^{(0)} + \beta f^{(1)} + \beta^2 f^{(2)} + \cdots \, ,
\label{eq:general_expansion}
\end{equation}
where
\begin{align}
f^{(0)} &= (e^{i\theta} - i) \int \frac{d\rho}{\sqrt{\rho}} K_{\frac{1}{2}}(M\rho)\, , \label{eq:p0} \\
f^{(1)} &= (i e^{i\theta} - 1) \int \frac{d\rho}{\sqrt{\rho}} \left[ D(M\rho) - \ln(a\rho) K_{\frac{1}{2}}(M\rho) \right]\, , \label{eq:p1} \\
f^{(2)} &= (e^{i\theta} - i) \int \frac{d\rho}{\sqrt{\rho}} \left[ -\frac{1}{2} D_2(M\rho) + \ln(a\rho) D(M\rho) - \frac{1}{2} \ln^2(a\rho) K_{\frac{1}{2}}(M\rho) \right]\, , \label{eq:p2}
\end{align}
with $D(z) = \left. \frac{\partial K_\nu}{\partial \nu} \right|_{\nu=\frac{1}{2}}$ and $D_2(z) = \left. \frac{\partial^2 K_\nu}{\partial \nu^2} \right|_{\nu=\frac{1}{2}}$

\subsection*{Case A: $M l \ll 1$} 

For $z = M\rho \ll 1$, we employ small-argument expansions \cite{DLMF}
\begin{align}
K_{\frac{1}{2}}(z) &= \sqrt{\frac{\pi}{2z}} e^{-z} \approx \sqrt{\frac{\pi}{2z}}\, , \label{eq:small_k} \\
D(z) &\approx -\sqrt{\frac{\pi}{2z}} \left( \ln z + \gamma \right)\, , \label{eq:small_d} \\
D_2(z) &\approx \sqrt{\frac{\pi}{2z}} \left( \ln^2 z + 2\gamma \ln z + \gamma^2 \right)\, , \label{eq:small_d2}
\end{align}
where $\gamma$ is the Euler-Mascheroni constant.
The zero-order term becomes
\begin{equation}
f^{(0)} \approx (e^{i\theta} - i) \sqrt{\frac{\pi}{2M}} \ln(a l) \, .
\label{eq:p0_result}
\end{equation}
The first-order term evaluates to
\begin{align}
f^{(1)} 
\approx -(i e^{i\theta} - 1) \sqrt{\frac{\pi}{2M}} \left[ \frac{1}{2} \ln^2(a l) + (\gamma + \ln M) \ln(a l) \right]\, .
\label{eq:p1_result}
\end{align}
The second-order term yields
\begin{equation}
f^{(2)} \approx -\frac{1}{2} (e^{i\theta} - i) \sqrt{\frac{\pi}{2M}} \left[ \frac{1}{3} \ln^3(a l) + (\gamma + \ln M) \ln^2(a l) \right]\, 
\label{eq:p2_result}
\end{equation}

\subsection*{Case C: $M/a \gg 1$} 

For $z = M\rho \gg 1$, we use large-argument expansions \cite{DLMF}:
\begin{align}
K_{\frac{1}{2}}(z) &= \sqrt{\frac{\pi}{2z}} e^{-z}\, , \label{eq:large_k} \\
D(z) &\sim \frac{1}{2} \sqrt{\frac{\pi}{2}} e^{-z} z^{-3/2}\, , \label{eq:large_d} \\
D_2(z) &\sim -\frac{3}{4} \sqrt{\frac{\pi}{2}} e^{-z} z^{-5/2}\, . \label{eq:large_d2}
\end{align}
The zero-order term becomes
\begin{equation}
f^{(0)} = (e^{i\theta} - i) \sqrt{\frac{\pi}{2M}} \int_{1/a}^{1/a + l} \frac{e^{-M\rho}}{\rho} d\rho \approx (e^{i\theta} - i) a \sqrt{\frac{\pi}{2}} M^{-3/2} e^{-M/a}\, .
\label{eq:p0_result_b}
\end{equation}
The first-order term evaluates to
\begin{equation}
f^{(1)} \approx \beta (i e^{i\theta} - 1) \sqrt{\frac{\pi}{2}} \frac{a^2}{2} M^{-5/2} e^{-M/a}\, .
\label{eq:p1_result_b}
\end{equation}
The second-order term yields
\begin{equation}
f^{(2)} \approx \beta^2 (e^{i\theta} - i) \sqrt{\frac{\pi}{2}} \frac{a^3}{8} M^{-7/2} e^{-M/a} \, .
\label{eq:p2_result_b}
\end{equation}

The dominant term depends on the regime:
For $M l \ll 1$: $f^{(0)} \sim \ln(a l)$ dominates. For $M/a \gg 1$: All terms are exponentially suppressed by $e^{-M/a}$.
The second-order terms are subdominant in both regimes.
Notably, all terms maintain dimensional consistency:
$[f] = [l \sqrt{a}] = [M^{-1}] \cdot [M^{1/2}] = [M^{-1/2}]$, and each term in the expansion has this dimension after accounting for the dimensionless factors.

These results demonstrate the transition from logarithmic behavior for light fields to exponential suppression for heavy fields, with the second-order terms in $\beta$ being subdominant in both cases.


\end{document}